\documentclass [aps, prb, reprint, twocolumn, floatfix, superscriptaddress]{revtex4}%
\usepackage{graphicx}
\usepackage{bm,amsmath,amssymb,mathrsfs,dcolumn}
\usepackage [colorlinks=true,linkcolor=blue,citecolor=blue,urlcolor=blue,linktoc=page,bookmarks=false,pdfstartview={FitH},pdfborder={0 0 0.0 [3 3]}]{hyperref}
\usepackage [usenames]{xcolor}
\hypersetup{pdfborder=0 0 0,colorlinks=true,citecolor=blue,linkcolor=blue}
\bibpunct{[}{]}{,}{n}{}{}

\def\be{\begin{equation}}
\def\ee{\end{equation}}
\def\ber{\begin{eqnarray}}
\def\eer{\end{eqnarray}}
\def\bs{\boldsymbol}
\begin{document}
\title{Spin-orbit coupling and linear crossings of dipolar magnons in  van der Waals antiferromagnets}
\author {Jie Liu}
\thanks{J. L. and L. W. contributed equally to this work.}
\affiliation{The Center for Advanced Quantum Studies and Department of Physics, Beijing Normal University, Beijing 100875, China}
\author{Lin Wang}
\thanks{J. L. and L. W. contributed equally to this work.}
\affiliation{Kavli Institute of Nanoscience, Delft University of Technology, P.O. Box 4056, 2600 GA Delft, The Netherlands}
\author {Ka Shen}
\email{kashen@bnu.edu.cn}
\affiliation{The Center for Advanced Quantum Studies and Department of Physics, Beijing Normal University, Beijing 100875, China}
\date{\today }


\begin{abstract}
  A magnon spin-orbit coupling, induced by the dipole-dipole interaction, is derived in monoclinic-stacked bilayer honeycomb spin lattice with perpendicular magnetic anisotropy and antiferromagnetic interlayer coupling. Linear crossings are predicted in the magnon spectrum around the band minimum in $\Gamma$ valley, as well as in the high frequency range around the zone boundary. The linear crossings in $K$ and $K'$ valleys, which connect the acoustic and optical bands, can be gapped when the intralayer dipole-dipole or Kitaev interactions exceed the interlayer dipole-dipole interaction, resulting in a phase transition from semimetal to  insulator. Our results are useful for analyzing the magnon spin dynamics and transport properties in van der Waals antiferromagnet.
\end{abstract}
\maketitle

\section{Introduction}

\begin{figure*}[t]
  \includegraphics[width=13cm]{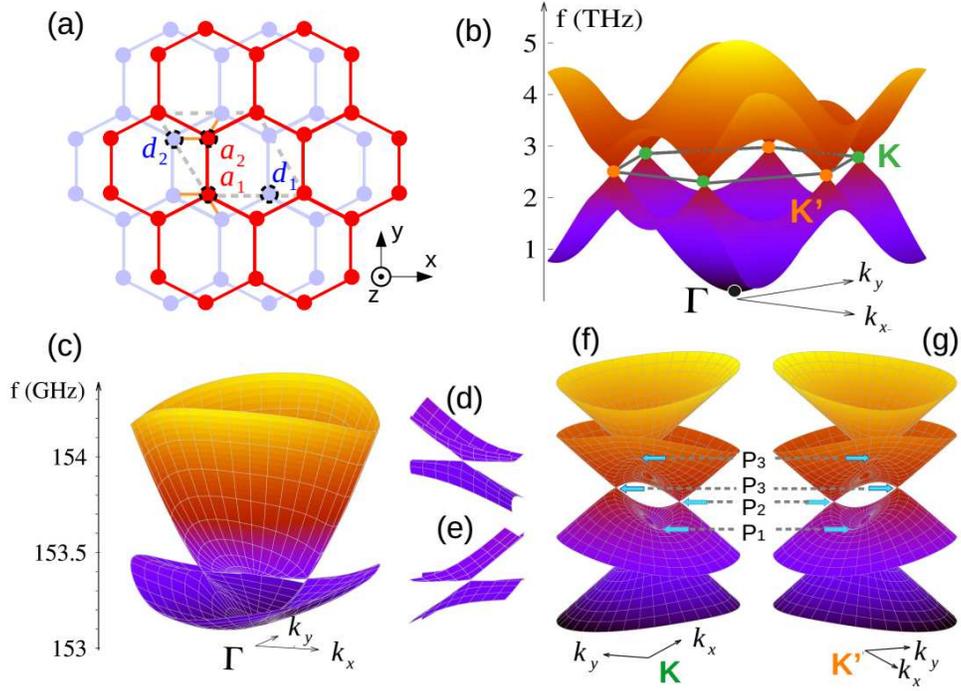}
  \caption{(a) The monoclinic lattice of bilayer CrI$_3$ with antiferromagnetic interlayer coupling. Only the magnetic atoms are shown. The coordinates of atoms within the unit cell are $\bs r_{a_1}=(0,0,\eta_z)a_0$, $\bs r_{a_2}=(0,1,\eta_z)a_0$, $\bs r_{d_1}=(2\sqrt{3}/3,0,-\eta_z)a_0$, and $\bs r_{d_2}=(-\sqrt{3}/3,1,-\eta_z)a_0$ with $a_0$ being the shortest distance between neighboring atoms. The interlayer distance is $2\eta_za_0$. The ground spin configuration corresponds to the spins in the upper (lower) layer colored in red (blue) orientating along $\hat z$ ($-\hat z$) direction. The orange bonds stand for those bearing antiferromagnetic interlayer coupling. (b) Full magnon spectrum and the fine structures in (c) $\Gamma$, (f) $K$, and (g) $K'$ valleys. (d) and (e) are the enlarged view around the left and right crossing points, respectively, in $\Gamma$ valley.}
  \label{lattice}
\end{figure*}

  Since the experimental demonstrations of magnetism in two dimensional (2D) van der Waals materials~\cite{xZhang17,xdXu17}, 2D magnetic materials and the spin excitations therein have attracted great research interests. For CrI$_3$, one of the most important 2D magnetic materials, the strong atomic magnetic anisotropy makes it beyond the Heisenberg model addressed in the Mermin-Wagner theorem~\cite{Mermin66} and is responsible for the existence of the long-range magnetic order.
Interestingly, the magnetic ground state of a bilayer CrI$_3$ is predicted to be either ferromagnetic or antiferromagnetic, depending on the way of stacking~\cite{Dixiao18,Jiwei19}. In particular, the monoclinic-stacked structure, as illustrated in Fig.~\ref{lattice}(a), has been demonstrated experimentally to be a $\cal PT$-symmetric antiferromagnet, where the two ferromagnetic monolayers align antiferromagnetically with a relative shift along the zigzag direction~\cite{SWWu19}.

The dynamics and transport properties of magnons, quanta of collective spin waves, have recently received intensive investigations in traditional bulk antiferromagnets~\cite{WangHL2014,Qiu16,Lebrun18,JShi20,Vaidya20}. One outstanding property of the magnons in antiferromagnets is the coexistence of different spin polarized modes, which supplies more interesting physics due to  the additional spin degree of freedom~\cite{Xiao17,Shen20}, compared to magnons in ferromagnets. The studies in van der Waals antiferromagnets in this direction, however, remain limited. In bilayer CrI$_3$, for instance, the magnetization dynamics of the uniform mode has been observed only very recently through ultrafast optical pump/magneto-optical Kerr probe technique~\cite{Zhang20} and magneto-Raman spectroscopy~\cite{Cenker20}. There is so far rare report on spin dynamic and transport of the propagating magnons. 

In this paper, we perform a theoretical study on the magnon spectrum of the monoclinic-stacked antiferromagnetic bilayer in Fig.~\ref{lattice}(a), by taking into account the magnetic anisotropy, exchange interaction and dipole-dipole interaction (DDI). Two linear crossings, as shown in Fig.~\ref{lattice}(c), are predicted around the band minimum near the $\Gamma$ point. Other linear crossing points in short wavelength regime with THz frequency, e.g., those in $K$ and $K'$ valleys shown Fig.~\ref{lattice}(f) and (g) and discussed in detail below, are also found.  The effective Hamiltonian, which captures the main features, is derived. Moreover, a phase transition between semimetal and insulator, will also be discussed.

\section{Model and Hamiltonian}
We model our spin system by a Hamiltonian including magnetic anisotropy, Zeeman term, exchange interaction and DDI
\ber
   {H}&=&\frac{K}{2}\sum_{i}(S_{i}^{z})^{2}+g\mu_{B}\sum_{i}\boldsymbol{S}_{i}\cdot\boldsymbol{B}-\sum_{\langle i,j \rangle}J_{ij}\boldsymbol{S}_{i}\cdot\boldsymbol{S}_{j}\nonumber\\
   &&\hspace{-1cm}\mbox{}+\frac{\mu_{0}(g\mu_{B})^{2}}{2}\sum_{i\ne j}\frac{R_{ij}^{2}(\boldsymbol{S}_{i}\cdot\boldsymbol{S}_{j})-3(\boldsymbol{R}_{ij}\cdot\boldsymbol{S}_{i})(\boldsymbol{R}_{ij}\cdot\boldsymbol{S}_{j})}{R_{ij}^{5}}.
   \label{Hami}
\eer
For a stable classical antiferromagnetic ground state indicated in Fig.~\ref{lattice}(a), we take the anisotropy parameter $K<0$ and the nearest intralayer and interlayer exchange parameters $J>0$ and $J'<0$, respectively. And the magnetic field is normal to the plane.

By applying the Holstein-Primakoff transformation~\cite{Holstein40},
\ber
S_{a}^{z}=S-a^{\dagger}a,& S_{a}^{+}=\sqrt{2S-a^{\dagger}a}a,\nonumber\\
S_{d}^{z}=-S+d^{\dagger}d,& S_{d}^{+}=d^{\dagger}\sqrt{2S-d^{\dagger}d},
\label{HP}
\eer
we derive the magnon Hamiltonian under the basis $(a_{1\boldsymbol{k}},a_{2\boldsymbol{k}},d_{1\boldsymbol{k}},d_{2\boldsymbol{k}},a_{1-\boldsymbol{k}}^{\dagger},a_{2-\boldsymbol{k}}^{\dagger},d_{1-\boldsymbol{k}}^{\dagger},d_{2-\boldsymbol{k}}^{\dagger})^{T}$ with $a_{i\bs k}$ ($d_{i\bs k}$) and $a^\dag_{i\bs k}$ ($d^\dag_{i\bs k}$) representing the magnon annihilation and creation operators for the $i$-th sublattice in the top (bottom) layer, respectively. The first line of Hamiltonian~(\ref{Hami}) leads to~\cite{Shen2019c,Shen20}
\be
H_{\boldsymbol{k},-\boldsymbol{k}}^0=
\left(\begin{array}{cccc}
[H_{a}^{ij}] & 0 & 0 & [\lambda_{\boldsymbol{k}}^{ij}]\\
0 & [H_{d}^{ij}] & [\lambda_{\boldsymbol{k}}^{ij}]^{\dagger} & 0\\
0 & [\lambda_{\boldsymbol{k}}^{ij}] & [H_{a}^{ij}] & 0\\
 {[\lambda_{\boldsymbol{k}}^{ij}]}^{\dagger} & 0 & 0 & [H_{d}^{ij}]
\end{array}\right),
\label{H0}
\ee
where each block is a $2\times 2$ matrix with $i,j=1,2$ and the diagonal ones read
\be
[H_{a(d)}^{ij}]=\left(\begin{array}{cc}
\epsilon_{a(d)} & \lambda_{\boldsymbol{k}}\\
\lambda_{\boldsymbol{k}}^{\ast} & \epsilon_{a(d)}
\end{array}\right).\label{Had}
\ee
The matrix elements are defined as
\ber
\epsilon_{a}&=&\omega_{{\rm ex}}+2\omega_{{\rm ex}}'+\omega_{{\rm an}}-\omega_{H},\\
\epsilon_{d}&=&\omega_{{\rm ex}}+2\omega_{{\rm ex}}'+\omega_{{\rm an}}+\omega_{H},\\
\lambda_{\boldsymbol{k}}&=&-\omega_{{\rm ex}}\gamma_{\boldsymbol{k}},\\
\lambda_{\boldsymbol{k}}^{ij}&=&\omega_{{\rm ex}}'\gamma_{\boldsymbol{k}}^{ij},
\eer
with $\omega_{{\rm ex}}=3SJ$, $\omega_{{\rm an}}=-KS$, $\omega_{{\rm ex}}'=-SJ'$, and $\omega_{H}=g\mu_{B}B$. The form factors are $\gamma_{\boldsymbol{k}}=\frac{1}{3}\sum_{i=1}^3e^{i\boldsymbol{k}\cdot\boldsymbol{\delta}_{i}}$ and $\gamma_{\boldsymbol{k}}^{ij}=e^{i\boldsymbol{k}\cdot\boldsymbol{\delta}_{ij}'}$ with the relative coordinates between the neighboring atoms being
\ber
\bs\delta_1&=&(0,a_0,0),\\
\bs\delta_2&=&(\frac{\sqrt{3}a_0}{2},-\frac{a_0}{2},0),\\
\bs\delta_3&=&(-\frac{\sqrt{3}a_0}{2},-\frac{a_0}{2},0),\\
\bs\delta_{11}'&=&\bs\delta_{22}'=(-\frac{\sqrt{3}a_0}{3},0,-2\eta_z a_0),\\
\bs\delta_{12}'&=&(-\frac{\sqrt{3}a_0}{6},\frac{a_0}{2},-2\eta_z a_0),\\
\bs\delta_{21}'&=&(\frac{\sqrt{3}a_0}{6},\frac{a_0}{2},-2\eta_z a_0).
\eer

The DDI term, i.e., the second line in Hamiltonian~(\ref{Hami}), gives
\be
H_{\boldsymbol{k},-\boldsymbol{k}}^{\rm DDI}=
\left(\begin{array}{cccc}
[A_{\boldsymbol{k}}^{a_{i}a_{j}}] & [B_{\boldsymbol{k}}^{d_{i}a_{j}}]^{\dagger} & [B_{\boldsymbol{k}}^{a_{i}a_{j}}]^{\dagger} & [A_{\boldsymbol{k}}^{d_{i}a_{j}}]^{\dagger}\\
{[B_{\boldsymbol{k}}^{d_{i}a_{j}}]} & [A_{\boldsymbol{k}}^{d_{i}d_{j}}] & [A_{\boldsymbol{k}}^{d_{i}a_{j}}] & [B_{\boldsymbol{k}}^{d_{i}d_{j}}]\\
{[ B_{\boldsymbol{k}}^{a_{i}a_{j}}]} & [A_{\boldsymbol{k}}^{d_{i}a_{j}}]^{\dagger} & [A_{\boldsymbol{k}}^{a_{i}a_{j}}] & [B_{\boldsymbol{k}}^{a_{i}d_{j}}]\\
{[A_{\boldsymbol{k}}^{d_{i}a_{j}}]} & [B_{\boldsymbol{k}}^{d_{i}d_{j}}]^{\dagger} & [B_{\boldsymbol{k}}^{a_{i}d_{j}}]^{\dagger} & [A_{\boldsymbol{k}}^{d_{i}d_{j}}]
\end{array}\right).
\label{DDI}
\ee
The matrix elements are defined as
\ber
A_{\boldsymbol{k}}^{\alpha\beta }&=&-\frac{S\mu_{0}(g\mu_{B})^{2}}{2} G^{\alpha\beta}_{\bs k},\\
B_{\boldsymbol{k}}^{\alpha\beta }&=&\frac{S\mu_{0}(g\mu_{B})^{2}}{2}F^{\alpha\beta}_{\bs k},
\eer
in which
  \ber
  F^{\alpha\beta}_{\bs k}&=&-3\sum_{mn}\frac{(X_{mn}^{\alpha\beta }-iY_{mn}^{\alpha\beta })^{2}}{(R_{mn}^{\alpha\beta})^{5}}e^{i{\bs k}\cdot\boldsymbol{R}_{mn}^{\alpha\beta}},\\
  G^{\alpha\beta}_{\bs k}&=&\sum_{mn}\frac{(R_{mn}^{\alpha\beta})^{2}-3(Z_{mn}^{\alpha\beta })^{2}}{(R_{mn}^{\alpha\beta})^{5}}e^{i{\bs k}\cdot\boldsymbol{R}_{mn}^{\alpha\beta}}.
\eer
with ${\bs R}_{mn}^{\alpha\beta}=m{\bs v_1}+n{\bs v_2}+{\bs r}_{\beta}-{\bs r}_{\alpha}$. The unit translation vectors read
\ber
    {\bs v_1}&=&(\sqrt 3 a_0,0,0),\\
    {\bs v_2}&=&(-\frac{\sqrt 3}{2} a_0,\frac{3}{2} a_0,0).
    \eer
    The entire magnon spectrum thus can be calculated from Hamiltonian (\ref{H0}) and (\ref{DDI}). The main features in the absence of the magnetic field are plotted in Fig.~\ref{lattice}, for which  we have adopted the parameters in bilayer CrI$_3$ with $S=3/2$, $K=-0.49$~meV, $J=2.2$~meV, $J'=-0.04$~meV, $a_0=3.98$~\AA, and $2\eta_za_0=3.98$~\AA~\cite{Dixiao18,LChen18}.

\begin{figure}[h]
  \includegraphics[width=5.8cm]{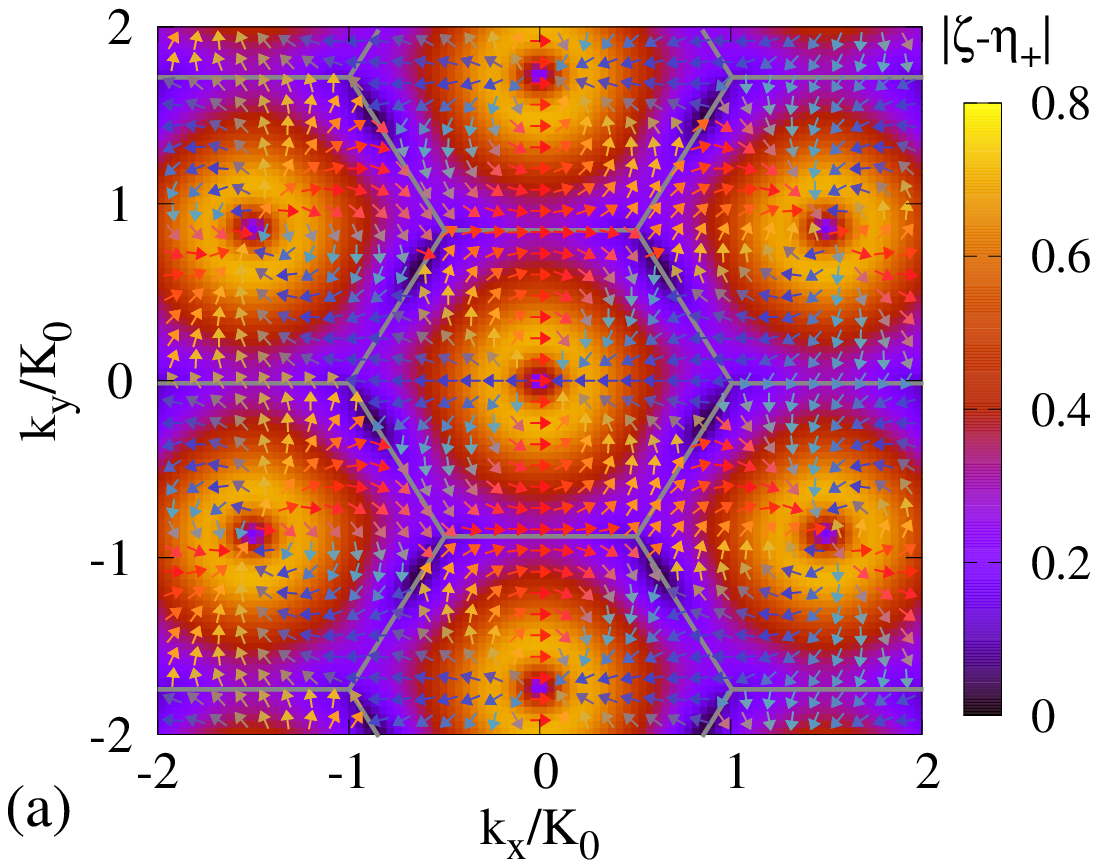}
  \includegraphics[width=5.8cm]{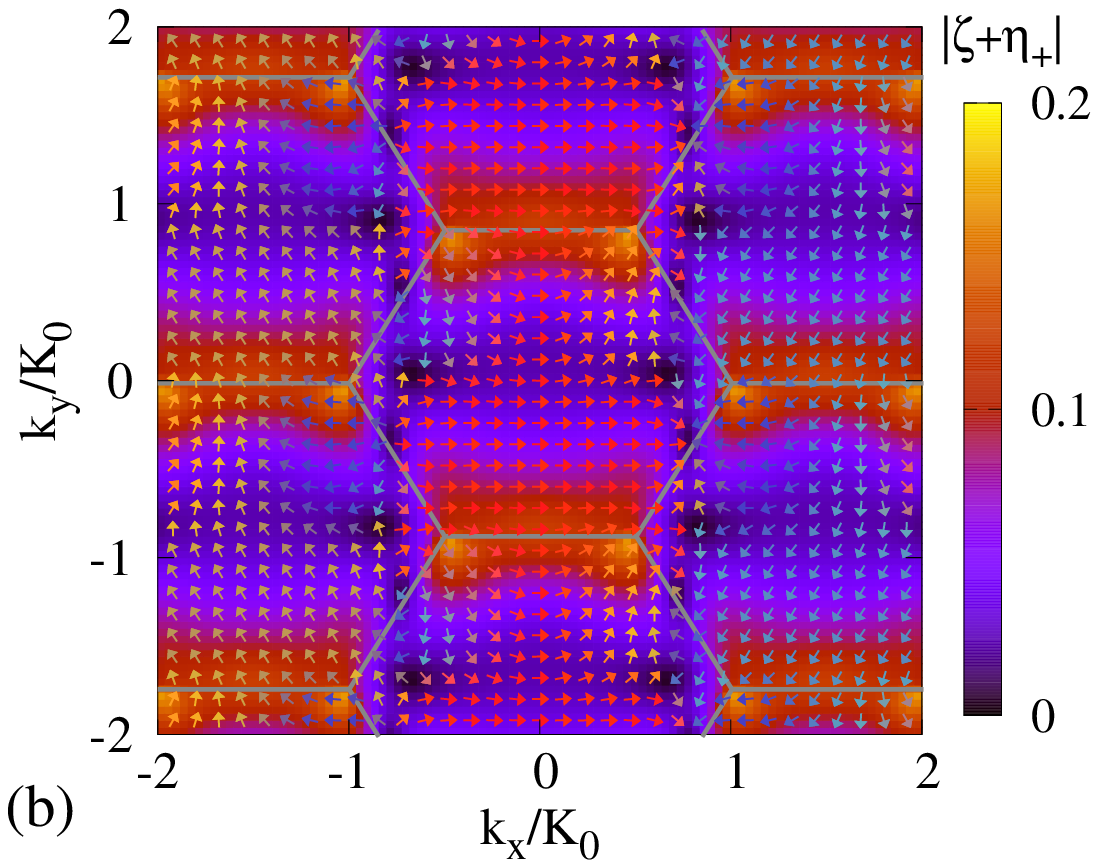}
  \includegraphics[width=5.8cm]{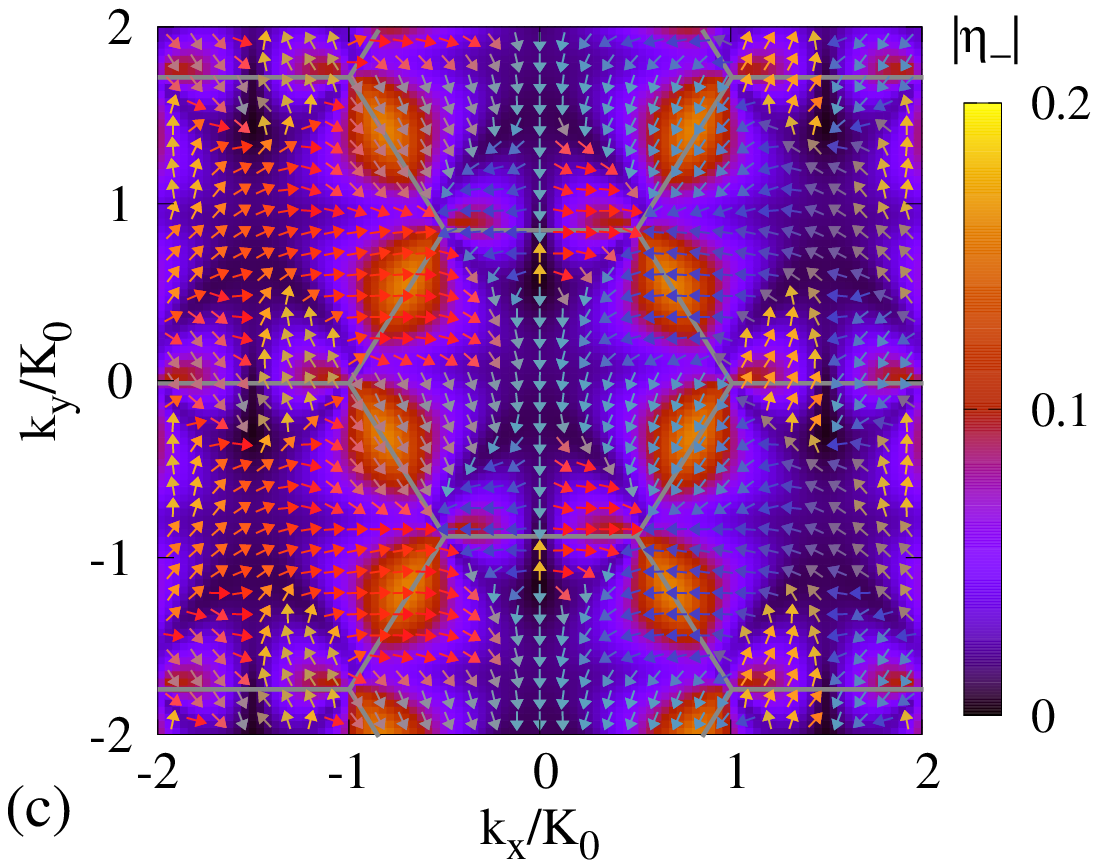}
  \caption{DDI-induced intraband spin-orbit parameters of (a) acoustic and (b) optical magnon bands and (c) the interband coupling. The arrows stand for the arguments of these complex parameters. $K_0={4\pi}/({3\sqrt{3} a_0})$.}
  \label{G_wave}
\end{figure}

\section{Magnon spin-orbit coupling}

\begin{figure}[t]
  \includegraphics[width=7cm]{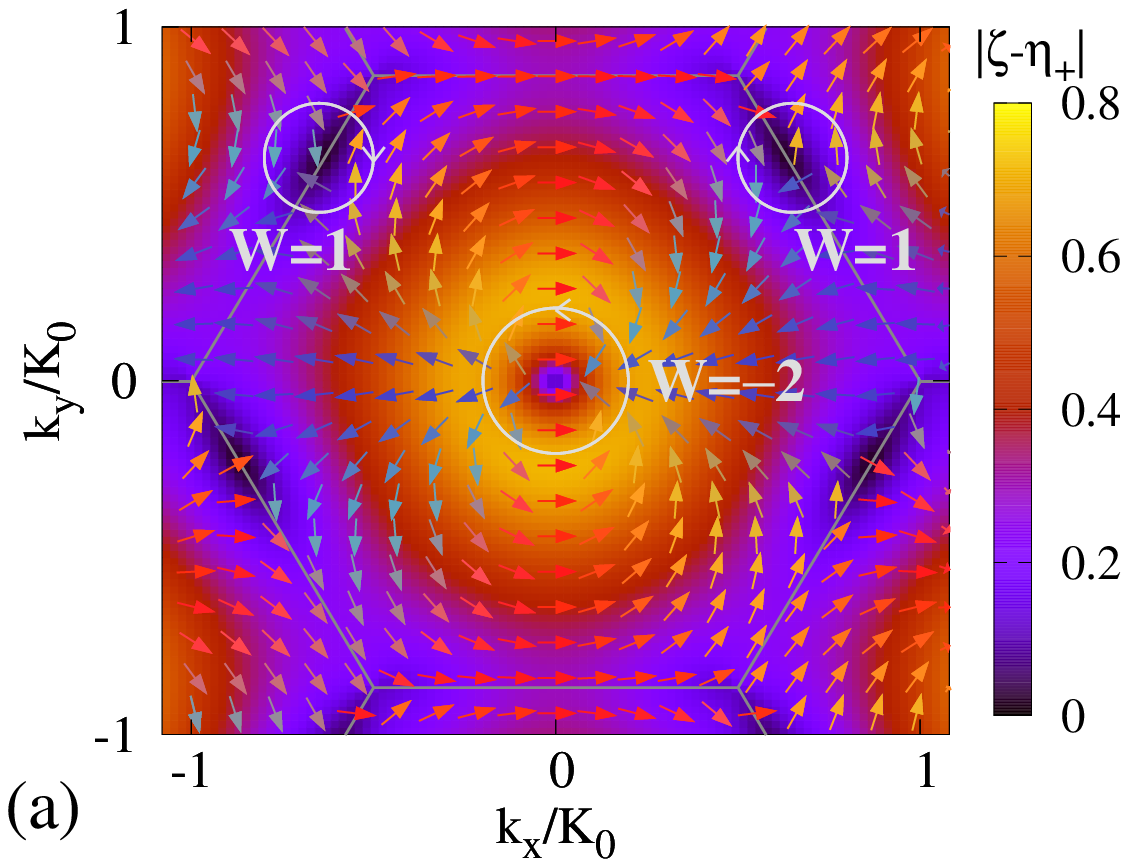}
  \includegraphics[width=7cm]{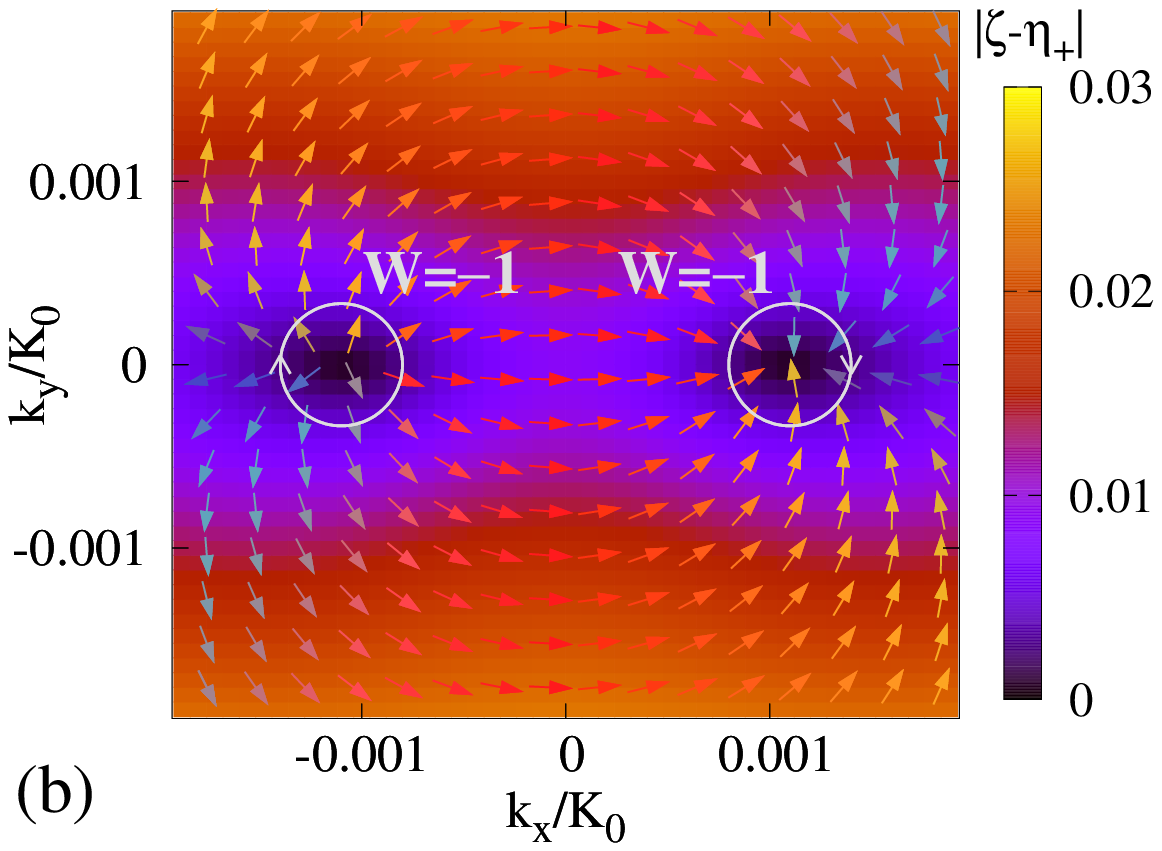}
  \caption{DDI-induced intraband spin-orbit field of the acoustic magnon bands (a) in first Brillouin zone and (b) in the vicinity of $\Gamma$ point. $W$ is the winding number of each loop. }
  \label{G_valley}
\end{figure}

For a better understanding of the numerical results, we perform an analytical analysis, for which we ignore the particle-hole coupling, i.e., the $4\times 4$ off-diagonal blocks in Hamiltonian (\ref{H0}) and (\ref{DDI}), by considering the fact that the interlayer exchange interaction ($\omega_{\rm ex}'\sim 0.06$~meV) and the DDI ($|A(B)_{\bs k}^{\alpha\beta}|\sim 0.01$~meV) in bilayer CrI$_3$ are much weaker than the particle-hole splitting (twice of the magnon gap) due to the anisotropy ($\omega_{\rm an}\sim 0.7$~meV) and intralayer exchange interaction ($\omega_{\rm ex}\sim 10$~meV), and therefore do not cause qualitative change in the magnon spectrum (unless otherwise clarified in Sect.~\ref{sec_phase}). Then, we can restrict our discussion within the particle subspace $(a_{1\boldsymbol{k}},a_{2\boldsymbol{k}},d_{1\boldsymbol{k}},d_{2\boldsymbol{k}})^T$. The reduced Hamiltonian can be in general written as
\be
H_{\bs k}=\left(\begin{array}{cccc}
\epsilon_{a}+\epsilon'_{\bs k} & \lambda_{\boldsymbol{k}}+\lambda_{\boldsymbol{k}}' & \zeta^{\ast}_{\bs k} & \eta_{\bs k}^{\ast}\\
\lambda_{\boldsymbol{k}}^{\ast}+\lambda_{\boldsymbol{k}}'^\ast & \epsilon_{a}+\epsilon'_{\bs k} & \eta_{\bs k}^{\prime\ast} & \zeta^{\ast}_{\bs k}\\
\zeta_{\bs k} & \eta_{\bs k}^\prime & \epsilon_{d}+\epsilon'_{\bs k} & \lambda_{\boldsymbol{k}}+\lambda_{\boldsymbol{k}}'\\
\eta_{\bs k} & \zeta_{\bs k} & \lambda_{\boldsymbol{k}}^{\ast}+\lambda_{\boldsymbol{k}}'^\ast & \epsilon_{d}+\epsilon'_{\bs k}
\end{array}\right),
\label{Hparticle}
\ee
with intralayer DDI parameters
\ber
\epsilon'_{\bs k}&=&A_{\bs k}^{a_1 a_1}=A_{\bs k}^{d_1 d_1},\\
\lambda'_{\bs k}&=&A_{\bs k}^{a_1 a_2}=A_{\bs k}^{d_1 d_2},
\eer
and interlayer ones
\ber
\zeta_{\bs k}&=&B_{\bs k}^{d_1 a_1},\\
\eta_{\bs k}&=&B_{\bs k}^{d_2 a_1}, \\
\eta^\prime_{\bs k}&=&B_{\bs k}^{d_1 a_2}.
\eer
It is convenient to transform Hamiltonian (\ref{Hparticle}) into the representation under the basis of the eigenstates of intralayer interaction, i.e., $(a^+_{\bs k},d^+_{\bs k},a^-_{\bs k},d^-_{\bs k})$ with
\be
a(d)^\pm_{\bs k}=\frac{1}{\sqrt 2}\left[a(d)_{1\bs k}\pm \frac{\lambda_{\bs k}^\ast+\lambda_{\bs k}^{\prime\ast}} {|\lambda_{\bs k}+\lambda_{\bs k}^\prime|}a(d)_{2\bs k} \right].
\ee
The Hamiltonian (\ref{Hparticle}) thus becomes
\be
\tilde H_{\bs k}=
\left(\begin{array}{cccc}
\epsilon^+_{a,\bs k} & \zeta^{\ast}_{\bs k}+\eta_{+,\bs k}^{\ast} & 0 & \eta_{-,\bs k}^{\ast}\\
\zeta_{\bs k}+\eta_{+,\bs k} & \epsilon^+_{d,\bs k} & -\eta_{-,\bs k} &0\\
0 & -\eta_{-,\bs k}^{\ast} & \epsilon^-_{a,\bs k} & \zeta^{\ast}_{\bs k}-\eta_{+,\bs k}^{\ast}\\
\eta_{-,\bs k} &0  & \zeta_{\bs k}-\eta_{+,\bs k} & \epsilon^-_{d,\bs k}
\end{array}\right),
\label{Htot}
\ee
with
\ber
\epsilon^{\pm}_{a(d),\bs k}&=&\epsilon_{a(d)}+\epsilon_{\bs k}^\prime \pm |\lambda_{\bs k}+\lambda_{\bs k}^\prime|,\\
\eta_{\pm,\bs k}&=&\frac{{(\lambda_{\boldsymbol{k}}^{\ast}+\lambda_{\boldsymbol{k}}^{\prime\ast})\eta'_{\bs k}\pm (\lambda_{\boldsymbol{k}}+\lambda_{\boldsymbol{k}}^{\prime})\eta_{\bs k}}}{{2|\lambda_{\boldsymbol{k}}+\lambda_{\boldsymbol{k}}^\prime|}}.
\eer
Here, the superscripts, ``$-$''  and ``$+$'', denote the acoustic and optical bands, respectively.  Since the excitations, $a_{\bs k}$ and $d_{\bs k}$, have opposite spin polarization~\cite{Shen20}, Hamiltonian (\ref{Htot}) reveals that the interlayer DDI introduces not only an intraband spin-orbit coupling to the acoustic and optical branches separately via $\zeta_{\bs k} -\eta_{+,\bs k}$ and $\zeta_{\bs k} +\eta_{+,\bs k}$, but also an interband spin-orbit coupling scaled by $\eta_{-,\bs k}$. The momentum dependence of these spin-orbit parameters are plotted in Fig.~\ref{G_wave}, in which inactive momentum points with vanishing value are observed. The situation of the acoustic branch, i.e., $\zeta_{\bs k} -\eta_{+,\bs k}$, is explicitly shown in Fig.~\ref{G_valley}. The vanishing value around the $\Gamma$ point and the boundary of Brillouin zone explains the formation of nodal points shown in Fig.~\ref{lattice}(c). The orientation of the spin-orbit field, as indicated by the colored arrows, varies with wave vector around the nodal points.

\subsection{Effective Hamiltonian and nodal points in $\Gamma$ valley}

In the vicinity of the $\Gamma$-point, $\lambda_{\bs k}\simeq -(1-k^2a_0^2/4)\omega_{\rm ex}$. The acoustic modes and the optical modes are well separated in frequency. Thanks to the relation $|\eta_{-,\bs k}|\ll \omega_{\rm ex}$, one can treat them separately. The Hamiltonian thus reduces to two subsystems involving only the spin degree of freedom
\be
\tilde H^\pm_{\bs k}=\left(\begin{array}{cc}
  \epsilon^\pm_{a,\bs k} & \zeta^{\ast}_{\bs k}\pm\eta_{+,\bs k}^{\ast} \\
\zeta_{\bs k}\pm\eta_{+,\bs k} & \epsilon^\pm_{d,\bs k}
\end{array}\right),
\ee
where the diagonal and off-diagonal components read
\ber
\epsilon^{\pm}_{a(d),\bs k}&=&\epsilon_{a(d)} -\frac{S\mu_0 (g\mu_B)^2}{2} (G^{a_1a_1}_{\bs k} \pm G^{a_1a_2}_{\bs k})\nonumber\\
&&\pm (1-\frac{a_0^2}4 k^2)\omega_{\rm ex},\\
\zeta_{\bs k}\pm\eta_{+,\bs k}&=&\frac{S\mu_0 (g\mu_B)^2}{2} (F_{\bs k}^{d_1 a_1} \mp \frac{F_{\bs k}^{d_1 a_2}+ F_{\bs k}^{d_2 a_1}}{2}).
\eer
After evalutating the summation in $F^{\alpha\beta}_{\bs k}$ and $G^{\alpha\beta}_{\bs k}$ in the long wavelength limit, as explained in Appendix~\ref{AppI} , we achieve an analytical expression
\be
\tilde H^\pm_{\bs k}
=\bar\epsilon^\pm_{\bs k} +(\omega_H\hat z+{\bs \Delta}^{\pm}_{\bs k})\cdot {\bs \sigma}.\label{Heff}
\ee
Here, the spin-independent energy can be expressed as
\ber
\bar\epsilon^{-}_{\bs k}&=&\omega_{\rm an}+2\omega_{\rm ex}'+f_z+f_z'+v_0k+\omega_{\rm ex}\frac{a_0^2}4 k^2,\\
\bar\epsilon^{+}_{\bs k}&=&\omega_{\rm an}+2(\omega_{\rm ex}+\omega_{\rm ex}')+f_z-f_z'-\omega_{\rm ex}\frac{a_0^2}4 k^2.
\eer
and the spin-orbit fields are
\ber
{\bs \Delta}^{-}_{\bs k}&=&(v_{0}k\cos2\phi_{\boldsymbol{k}}-f_{0},v_{0}k\sin2\phi_{\boldsymbol{k}},0),\label{soc_acoustic}\\
{\bs \Delta}^{+}_{\bs k}&=&(-3f_{0},0,0),
\eer
where $f_0$ and $v_0$ are positive real numbers. It is interesting to notice that while the spin-orbit coupling of the optical band is a simple constant, the one of the acoustic band contains an additional contribution varying with the direction of momentum. Such an angular dependence is a common feature of dipolar field in the long wavelength limit~\cite{Shen20}. Another important feature one can observe from spin-orbit field~(\ref{soc_acoustic}) is that, distinct from the three dimensional (3D) case~\cite{Shen20}, the magnitude of the angular dependent term here is linear in $k$.

Apparently, the spin-orbit field (\ref{soc_acoustic}) vanishes at ${\bs k}_\pm=(\pm f_0/v_0,0)\simeq (\pm 0.001 K_0,0)$, which well explains the numerical results in Fig.~\ref{G_valley}(b). At $\omega_H=0$, ${\bs k}_\pm$ corresponds to the two nodal points in Fig.~\ref{lattice}(c). We expand the spin-orbit coupling (\ref{soc_acoustic}) nearby ${\bs k}={\bs k}_{\pm}+\tilde{\bs k}$ and obtain an effective Hamiltonian up to the linear order in $\tilde {\bs k}$
\ber
H^-_{\pm}(\tilde{{\bs k}})&=&\pm (v_{0}{\tilde k_x}\sigma_x+ 2v_{0}{\tilde k_y}\sigma_y)\nonumber\\
&=&\left(\begin{array}{cc}
  0& \pm v_0 (\tilde k_x -2 i\tilde k_y)\\
  \pm v_0 (\tilde k_x +2 i\tilde k_y) &0
  \end{array}\right), \label{Hpmx}
\eer
where the spin-independent term $\bar \epsilon^-_{\bs k_{\pm}}$ has been discarded. This Hamiltonian has chiral symmetry $CH^-_{\pm}({\tilde {\bs k}})C^{-1}=-H^-_{\pm}(\tilde{{\bs k}})$ with $C=\sigma_z$. As a result, the chirality of 
the nodal points is characterized by the winding number
\be
W=({1}/{2\pi i})\oint_{\rm L}{d\xi({\tilde{\bs k}})}/{\xi({\tilde{\bs k}})}, \label{eqWinding}
\ee
where the integration performed over a closed loop around the nodal point and $\xi_{\pm}({\tilde{\bs k}})$ is defined as the phase factor of the off-diagonal matrix element in (\ref{Hpmx}), i.e., 
\be
\xi_{\pm}({\tilde{\bs k}})=\pm\frac{{\tilde k_x}-2i{\tilde k_y}}{|{\tilde k_x}-2i{\tilde k_y}|}.
\ee
This leads to chirality of $-1$ for both nodal points. The non-zero chirality, the linear crossing, and the two-fold degeneracy together indicate that these nodal points can be regarded as a magnon analogue of the 2D Weyl points recently proposed in electronic system~\cite{GSu19}.

As shown in Fig.~\ref{G_valley}(b), the winding number $-1$ also coincides with the direct observation of a $2\pi$-rotation of the spin-orbit field through a closed loop around each nodal point. The total winding number of the $\Gamma$ valley [for a single loop besieging both nodal points in Fig.~\ref{G_valley}(a)] is therefore $-2$, being the same as the one around the dipolar-induced nodal line in 3D cubic lattice~\cite{JLiu20}. Figure~\ref{G_valley}(a) shows that another two nodal points at the zone boundary both have winding number $+1$, compensating the chirality from ${\bs k}_\pm$ near $\Gamma$ valley. In contrast to the 3D Weyl points, which are robust against any perturbation~\cite{Wan11,Burkov11,Hasan15,HDing15}, the 2D Weyl points can be gapped by particular perturbation~\cite{GSu19}. In the present case, for example, the inclusion of a non-vanishing $\omega_H$ due to an out-of-plane magnetic field, according to the effective Hamiltonian (\ref{Heff}),  opens a gap at the crossing points, very similar to the situation in 2D electron gas with Dresselhaus- or Rashba-type in-plane spin-orbit field. 

\subsection{Nodal points in $K$ and $K'$ valley}

At $K$ and $K'$ points, the band splitting $|\lambda_{\bs k}+\lambda_{\bs k}^\prime|$ vanishes, therefore, one has to treat the acoustic and optical bands together by using the complete $4\times 4$ Hamiltonian~(\ref{Htot}). By expanding the Hamiltonian around these points
\be
\bs k={\bs K(\bs K')}+q(\cos\theta_{\bs q},\sin\theta_{\bs q})
\ee
with ${\bs K} (\bs K')=(\mp K_0,0)$ and $q\ll K_0$, we obtain
\ber
\eta_{\pm,\bs K+\bs q} &\simeq &-\zeta_{\bs K}[e^{-i\theta_{\bs q}}\pm e^{i(\theta_{\bs q}-2\pi/3)}]/2,\label{etaK}\\
\eta_{\pm,\bs K' +\bs q}&\simeq &\pm \zeta_{\bs K'}[e^{-i\theta_{\bs q}}\pm e^{i(\theta_{\bs q}+2\pi/3)}]/2,\label{etaKm}
\eer
with $\zeta_{\bs K'}=\zeta_{\bs K}^\ast$. The derivation of Eqs.~(\ref{etaK}) and (\ref{etaKm}) is given in Appendix~\ref{socKKm}. At $\theta_{\bs q}=\pm \pi/3$, all elements of the off-diagonal blocks depending solely on $\eta_{-,\bs k}$ vanish for $K$ and $K'$ valleys, respectively. The four dispersion curves along this momentum line become linear in $q$, i.e., $\bar \epsilon_{\bs K}-v_F q\pm \sqrt {3}|\zeta_{\bs K}|$ and $\bar \epsilon_{\bs K}+v_F q\pm |\zeta_{\bs K}|$. The intersections between them give rise to four nodal points in each valley, namely,
\ber
{\bs k}_{1,\tau}&=&\frac{(\sqrt{3}+1)|\zeta_{\bs K}|}{2v_F}(\frac{\tau}2,\frac{\sqrt{3}}2),\\ 
{\bs k}_{2,\tau}&=&\frac{(\sqrt{3}-1)|\zeta_{\bs K}|}{2v_F}(\frac{\tau}2,\frac{\sqrt{3}}2),\\
{\bs k}_{3,\tau}&=&\frac{(\sqrt{3}+1)|\zeta_{\bs K}|}{2v_F}(-\frac{\tau}2,-\frac{\sqrt{3}}2),\\
{\bs k}_{4,\tau}&=&\frac{(\sqrt{3}-1)|\zeta_{\bs K}|}{2v_F}(-\frac{\tau}2,-\frac{\sqrt{3}}2).
\eer
Here, $\tau$ is the valley index with $K$ $(\tau=1)$ and $K'$ $(\tau=-1)$. This is consistent with the spectrum from a full calculation plotted in Figs.~\ref{lattice}(f) and (g).
To uncover the nature of these nodal points, we again expand the Hamiltonian nearby ${\bs k}={\bs k}_{i,\tau}+\tilde{\bs k}$ and derive a $2\times 2$ effective Hamiltonian for each nodal point under the basis of the two branches involving in as
\ber
H_{1,\tau}&=&n_{z,\tau}s_z-n_{\|,\tau}(\tau s_x +s_y),\\
H_{2,\tau}&=&n_{z,\tau}s_z+n_{\|,\tau}(\tau s_x-s_y),\\
H_{3,\tau}&=&-n_{z,\tau}s_z-n_{\|,\tau}(\tau s_x-s_y),\\
H_{4,\tau}&=&-n_{z,\tau}s_z+n_{\|,\tau}(\tau s_x+s_y),
\eer
where
\ber
n_{z,\tau}&=&({v_F}/{2})(\tilde k_{x}\tau+\sqrt{3}\tilde k_{y}),\\
n_{\|,\tau}&=&({v_F}/{4})(\sqrt{3}\tilde k_{x}\tau-\tilde k_{y}).
\eer
Here, $s_{i=x,y,z}$ are Pauli matrices. 
These effective Hamiltonians can be classified into two groups, i.e.,
\be
H_{\pm}=n_zs_z+n_{\parallel}(s_x\pm s_y),
\ee
which have chiral symmetry
\be
C_{\pm}=(s_x\mp s_y)/\sqrt{2},
\ee
and therefore can be brought to a block off-diagonal form by a unitary transformation $U_{\pm}$. 
\ber
U_{\pm}^{-1}H_{\pm}U_{\pm}=
\left(\begin{array}{cc}
0 & -n_z \mp i\sqrt{2}n_{\parallel}\\
-n_z\pm i\sqrt{2}n_{\parallel} & 0
\end{array}\right),   \label{HII}
\eer
with 
\ber
U_{\pm}=\frac{1}{2}
\left(\begin{array}{cc}
1\pm i & -1\mp i\\
\sqrt 2 & \sqrt 2
\end{array}\right).   
\eer
By substituting the phase factor of the off-diagonal matrix element in Hamiltonian (\ref{HII}), i.e.,
\ber
\xi_{\pm}=-\frac{n_z \pm i\sqrt{2}n_{\parallel}}{|n_z \pm i\sqrt{2}n_{\parallel}|}
\eer
into Eq.~(\ref{eqWinding}), we find that all four nodal points near $K$ ($K'$) valley have the same chirality of $+1$ $(-1)$. By considering their two-fold degeneracy and the linear dispersion nearby, we conclude that these nodal points are also 2D Weyl points~\cite{GSu19}.

\begin{figure}[h]
  \includegraphics[width=7cm]{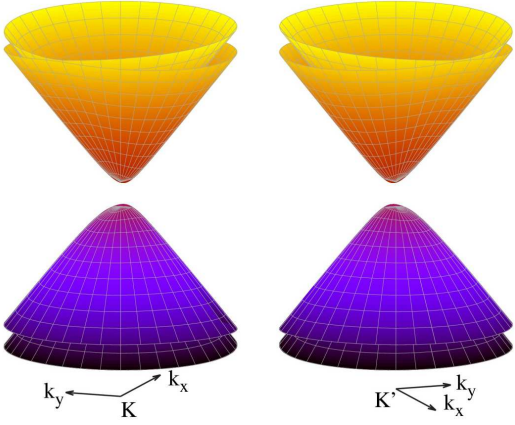}
  \caption{Magnon spectrum around the K and K' points from the calculation, in which the interlayer DDI is suppressed by increasing the interlayer distance by $30\%$.}
  \label{TI}
\end{figure}

\subsection{Phase transition from semimetal and insulator}\label{sec_phase}

We should point out that the particle-hole coupling due to the intralayer DDI, which has been neglected in the above analysis, actually is able to cause quantitative or even qualitative change for different parameter choices. Specifically, the nodal points $P_{1{\text -}4}$ in Figs.~\ref{lattice}(f) and (g) from full calculation are found to deviate from the momentum line with $\theta_{\bs q}=\pm\pi/3$. By increasing the ratio between the strengths of the intra- and inter-layer DDI via decreasing the interlayer distance,  $P_2$ and $P_3$ become closer and finally overlap at a critical ratio. When this ratio increases further, a global gap is opened between the acoustic and optical branches, leading to a transition from semimetal phase to insulating phase. The fine structure around K and K' points for the gapped phase is plotted in Fig.~\ref{TI}, for which the interlayer distance in enhanced by $30\%$. 

  According to Eq.~(\ref{Ba1a2K}), the particle-hole coupling induced by intralayer DDI gives
  \be
  B_{\boldsymbol{K}(\boldsymbol{K}')}^{a_{1}a_{2}}=B_{\boldsymbol{K}'(\boldsymbol{K})}^{a_{2}a_{1}}
  =f_K [1+e^{i\frac{2\pi}{3}(1\pm1)}+e^{-i\frac{2\pi}{3}(1\pm1)}],
  \label{Ba1a2}
  \ee
  which suggests $B_{\boldsymbol{K}}^{a_{1}a_{2}}=B_{\boldsymbol{K}'}^{a_{2}a_{1}}=0$ and $B_{\boldsymbol{K}}^{a_{2}a_{1}}=B_{\boldsymbol{K}'}^{a_{1}a_{2}}\ne 0$. Similarly, one has $B_{\boldsymbol{K}}^{d_{1}d_{2}}=B_{\boldsymbol{K}'}^{d_{2}d_{1}}=0$ and $B_{\boldsymbol{K}}^{d_{2}d_{1}}=B_{\boldsymbol{K}'}^{d_{1}d_{2}}\ne0$.

On the other hand, as recently shown in Ref.~\cite{Aguilera20}, the Kitaev interaction, an anisotropic term of the intralayer exchange interaction between the nearest neighboring Cr atoms, can produce a similar insulating phase in single layer CrI$_3$. To examine its role in our bilayer structure, we take into account this additional intralayer interaction~\cite{Kitaev06,HXiang18}
\be
H^{\rm K}=-K^A\sum_{\langle i,j\rangle} ({\bs S}_i \cdot \hat{\bs p}_{{\bs R}_{ij}})\cdot ({\bs S}_j \cdot \hat{\bs p}_{{\bs R}_{ij}}),
\ee
where
\ber
\hat{\bs p}_{{\bs \delta}_1}&=&\left(-\sqrt{\frac{2}{3}},0,\sqrt{\frac{1}{3}}\right),\\
\hat{\bs p}_{{\bs \delta}_2}&=&\left(\sqrt{\frac{1}{6}},\sqrt{\frac{1}{2}},\sqrt{\frac{1}{3}}\right),\\
\hat{\bs p}_{{\bs \delta}_3}&=&\left(\sqrt{\frac{1}{6}},-\sqrt{\frac{1}{2}},\sqrt{\frac{1}{3}}\right).
\eer
After applying the aforementioned standard procedures, we obtain 
\be
H^{K}=\left(\begin{array}{cccc}
  [H_{\boldsymbol{k}}^{\prime ij}] & 0 & [{\cal K}_{\boldsymbol{k}}^{ij}]^\dagger & 0\\
0 & [H_{\boldsymbol{k}}^{\prime ij}] & 0 & [{\cal K}_{\boldsymbol{k}}^{ij}]\\{}
[{\cal K}_{\boldsymbol{k}}^{ij}] & 0 & [H_{\boldsymbol{k}}^{\prime ij}] & 0\\
0 & [{\cal K}_{\boldsymbol{k}}^{ij}]^{\dagger} & 0 & [H_{\boldsymbol{k}}^{\prime ij}]
\end{array}\right),
\ee
where the non-vanishing blocks are expressed as
\ber
[H_{\boldsymbol{k}}^{\prime ij}]&=&K^{A}S\left(\begin{array}{cc}
1 & -\gamma_{\boldsymbol{k}}\\
-\gamma_{\boldsymbol{k}}^{\ast} & 1
\end{array}\right),\\
{[{\cal K}_{\boldsymbol{k}}^{ij}]}&=&K^{A}S\left(\begin{array}{cc}
0 & -\tilde{\gamma}_{\boldsymbol{k}}\\
-\tilde{\gamma}_{-\boldsymbol{k}} & 0
\end{array}\right).
\eer
The form factor $\gamma_{\bs k}$ here is the same as above and
\be
\tilde{\gamma}_{\boldsymbol{k}}=\frac{1}{3}(e^{i\boldsymbol{k}\cdot\boldsymbol{\delta}_{1}}+e^{i2\pi/3+i\boldsymbol{k}\cdot\boldsymbol{\delta}_{2}}+e^{-i2\pi/3+i\boldsymbol{k}\cdot\boldsymbol{\delta}_{3}}).
\ee
As one may notice, the diagonal blocks can be included into the Hamiltonian (\ref{Had}) by simply replacing the exchange parameter $\omega_{\rm ex}$ by $\omega_{\rm ex}+K^AS$. The off-diagonal blocks is additive to those from intralayer DDI $[B_{\bs k}^{a_i a_j}]$ and $[B_{\bs k}^{d_i d_j}]$. In particular, at $K$ and $K'$ points, we have
\be
\tilde{\gamma}_{\boldsymbol{K}(\boldsymbol{K}')}=\frac{1}{3}[1+e^{i\frac{2\pi}{3}(1\mp1)}+e^{-i\frac{2\pi}{3}(1\mp1)}],
\ee
which contains the same factor as Eq.~(\ref{Ba1a2}) and gives ${\cal K}_{\bs K}^{21}={\cal K}_{\bs K'}^{12}=0$, but ${\cal K}_{\bs K}^{12}={\cal K}_{\bs K'}^{21}\ne 0$. This indicates that the Kitaev affects the spectrum of  K and K' valleys in the same way as the intralayer DDI.

Therefore, a material with weaker Kitaev interaction is preferred for experimental observation of $K$($K'$)-valley nodal points. Another option would be to use an artificial structure to avoid anisotropic exchange interaction. Nevertheless, the linear crossings between the two spin bands in the $\Gamma$-valley are robust against the intralyer DDI and Kitaev interaction, even in the insulating phase. 

\subsection{Discussion on ${\cal PT}$-symmetry}
Before closing this paper, we would like to discuss the role of the ${\cal PT}$-symmetry in magnonic system. It is well known that in a $\cal PT$-symmetric fermionic system, Weyl fermions are forbidden because the $\cal PT$-symmetry introduces two-fold degeneracy of Weyl cones~\cite{PXTang16}. In the present magnonic case, such a degeneracy is removed by the interlayer DDI, which can 
be understood as follows. Without any interlayer coupling, for any magnon mode in the top layer $a_{\bs k}$, one can find its $\cal PT$ partner in the bottom layer $b_{\bs k}$. The bosonic nature of magnons requires $({\cal PT})^2=1$, and therefore ${\cal PT}a_{\boldsymbol{k}}=b_{\bs k}$ and ${\cal PT}b_{\boldsymbol{k}}=a_{\boldsymbol{k}}$. The interlayer DDI then introduces a coupling between 
$a_{\boldsymbol{k}}$ and $b_{\boldsymbol{k}}$ and generates hybrid eigenstates generally in form of
\be
\psi_{\bs k}=\frac{1}{\sqrt{2}}(a_{\boldsymbol{k}}+e^{i\delta_{\bs k}}b_{\boldsymbol{k}}).\label{psik}
\ee
The phase factor $\delta_{\bs k}$ relies on the explicit expression  of the coupling. The $\cal PT$ partner of $\psi_{\bs k}$ reads
\be
{\cal PT} \psi_{\bs k}=e^{-i\delta_{\bs k}}\frac{1}{\sqrt{2}}(a_{\boldsymbol{k}}+e^{i\delta_{\bs k}}b_{\boldsymbol{k}}),
\ee
equivalent to $\psi_{\bs k}$ except a marginal global phase factor $e^{-i\delta_{\bs k}}$. In other words, $\psi_{\bs k}$ itself is $\cal PT$-symmetric and no additional degeneracy is necessary. By contrast, the relation $({\cal PT})^{2}=-1$ in fermionic systems results in the $\cal PT$ partner of a hybrid state (\ref{psik}) as
\be
   {\cal PT} \psi_{\bs k}=e^{-i\delta_{\bs k}}\frac{1}{\sqrt{2}}(-a_{\boldsymbol{k}}+e^{i\delta_{\bs k}}b_{\boldsymbol{k}}),
\ee
which is orthogonal to $\psi_{\bs k}$, indicating that ${\cal PT}\psi_{\bs k}$ and $\psi_{\bs k}$ must be different states.

\section{Summary}
In summary, we predict a magnon spin-orbit coupling due to dipole-dipole interaction in monoclinic-stacked van der Waals antiferromagnetic bilayer. Such a spin-orbit coupling is expected to activate the intrinsic magnon spin relaxation mechanism and magnon spin Hall effect recently predicted in Ref.~\cite{Shen20}. Specifically, in the long wavelength limit, the spin-orbit coupling contains both momentum-independent and linearly momentum-dependent effective magnetic fields, which give rise to  two nodal points. Due to their low energy, these magnon states would have a large thermal population and are relevant even at low temperature. The sub-THz range of their frequencies, although much higher than the usual ferromagnetic resonance, is already achievable by current techniques~\cite{JShi20,Vaidya20,Zhang20,Cenker20}, which supports their observation and potential applications in magnonics. 
In the $K$ and $K'$ valleys, four nodal points are found in each valley. These nodal points connect the acoustic and optical magnon bands and  make a magnonic semimetal. By tuning the interlayer distance or introducing intralayer Kitaev interaction, a phase transition to insulating phase is predicted.

\begin{acknowledgments}
  This work is supported by the National Natural Science Foundation of China (Grants No.11974047), the Fundamental Research Funds for the Central Universities (Grant No. 2018EYT02) and the Netherlands Organisation for Scientific Research (NWO/OCW), as part of the Frontiers of Nanoscience program.

\end{acknowledgments}

\appendix
\section{Dipolar interaction in the long wavelength limit}\label{AppI}
We now calculate the summation appearing in the interlayer DDI
\be
F^{\alpha\beta}_{\bs k}=-3\sum_{mn}\frac{(X_{mn}^{\alpha\beta }-iY_{mn}^{\alpha\beta })^{2}}{(R_{mn}^{\alpha\beta})^{5}}e^{i{\bs k}\cdot\boldsymbol{R}_{mn}^{\alpha\beta}}.\label{Fab}
\ee
Focusing on the long wavelength regime, one can take a cut-off distance $\rho$ satisfying $1/k\gg\rho\gg a_{0}$. For all the in-plane distance shorter than $\rho$, it is safe to use $e^{i{\bs k}\cdot\boldsymbol{R}_{mn}^{\alpha\beta}}\simeq 1$.

\ber
F^{\alpha\beta}(\bs k)
&\simeq&-3\sum_{|\bs k\cdot \bs R^{\alpha\beta}_{mn}|>k\rho}\frac{(X_{mn}^{\alpha\beta }-iY_{mn}^{\alpha\beta })^{2}}{(R_{mn}^{\alpha\beta})^{5}}e^{i{\bs k}\cdot\boldsymbol{R}_{mn}^{\alpha\beta}}\nonumber\\
&&	-3\sum_{|\bs k \cdot \bs R_{mn}^{\alpha\beta}|<k\rho}\frac{(X_{mn}^{\alpha\beta }-iY_{mn}^{\alpha\beta })^{2}}{(R_{mn}^{\alpha\beta})^{5}}\nonumber\\
&=&-\frac{1}{A}\int_{r>\rho}d\boldsymbol{r}e^{i{\bs k}\cdot\boldsymbol{r}}(\partial_{x}^{2}-\partial_{y}^{2}-2i\partial_{x}\partial_{y})\frac{1}{\sqrt {r^2+h^2}}\nonumber\\
&&	-\frac{1}{A}(\beta^{\alpha\beta}_{xx}-\beta^{\alpha\beta}_{yy}-2i\beta^{\alpha\beta}_{xy}),
\label{fkb}
\eer
with $A$ being the area of a unit cell and $h$ the interlayer distance. Apparently, the atomistic detail of a specific crystal only affects the $\bs k$-independent parameters $\beta_{ij}^{\alpha\beta}$. The $\bs k$-dependent term can be calculated analytically.

By applying partition integration, one obtains
\ber
&&\int_{r>\rho}dxdye^{i{\bs k}\cdot\boldsymbol{r}}\partial_{y}^2\frac{1}{\sqrt {r^2+h^2}}\nonumber\\
&=&-(\oint_{L}dxe^{i{\bs k}\cdot\boldsymbol{r}}\frac{y}{({r^2+h^2})^{3}}
-\oint_{r=\rho}dxe^{i{\bs k}\cdot\boldsymbol{r}}\frac{y}{({r^2+h^2})^{3/2}})
\nonumber\\
&&+ik_{y}\oint_{L}dxe^{i{\bs k}\cdot\boldsymbol{r}}\frac{1}{\sqrt {r^2+h^2}}-ik_{y}\oint_{r=\rho}dxe^{i{\bs k}\cdot\boldsymbol{r}}\frac{1}{\sqrt {r^2+h^2}}\nonumber \\
&&
-k_{y}^{2}\int_{r>\rho}dxdye^{i{\bs k}\cdot\boldsymbol{r}}\frac{1}{\sqrt {r^2+h^2}}.
\label{eqyy}
\eer
Here, $L$ represents the outer boundary of the entire 2D system, which is assumed be to sufficiently large so that the factor $e^{-i\boldsymbol{k}\cdot\boldsymbol{r}}$ oscillates at the boundary, resulting in a significant reduction of the integration over $L$. At the inner surface with $r=\rho$ , we have $e^{-i\boldsymbol{k}\cdot\boldsymbol{r}}\simeq1$. Considering the interlayer distance $h$ of the same order of $a_0$, we have $h\ll \rho$ and therefore $\sqrt{r^2+h^2}\approx r$ for $r\ge \rho$. Equation~(\ref{eqyy}) then gives
\ber
&&\int_{r>\rho}dxdye^{i{\bs k}\cdot\boldsymbol{r}}\partial_{y}^{2}\frac{1}{\sqrt{r^2+h^2}}\nonumber\\
&\simeq&\frac{2\pi\rho^{2}}{(\rho^{2}+h^{2})^{3/2}}-\frac{k_{y}^{2}}{k}\int_{0}^{2\pi}d\theta\int_{k\rho}^{\infty}d\xi e^{i\xi\cos\theta}\nonumber\\
&=&	\frac{2\pi\rho^{2}}{(\rho^{2}+h^{2})^{3/2}}-2\pi\frac{k_{y}^{2}}{k}.
\eer
Similarly, one can calculate the other two integrals in Eq.~(\ref{fkb})
\ber
\int_{r>\rho}dxdye^{i{\bs k}\cdot\boldsymbol{r}}\partial_{x}^2\frac{1}{\sqrt{r^2+h^2}}
&=&\frac{2\pi\rho^{2}}{(\rho^{2}+h^{2})^{3/2}}-2\pi\frac{k_{x}^{2}}{k},\nonumber\\
\\
\int_{r>\rho}dxdye^{i{\bs k}\cdot\boldsymbol{r}}\partial_{y}\partial_{x}\frac{1}{\sqrt{r^2+h^2}}
&=&-2\pi\frac{k_{y}k_{y}}{k}.
\eer
Therefore, the summation (\ref{Fab}) can be expressed as
\be
F^{\alpha\beta}_{\bs k}=\frac{1}{A}(\beta_{yy}^{\alpha\beta}-\beta_{xx}^{\alpha\beta}+2i\beta_{xy}^{\alpha\beta}+2\pi ke^{-2i\theta_{\boldsymbol{k}}}),
\ee
with $\theta_{\bf k}$ representing the angle of $\bs k$ with respect to $x$-axis.

The spin-orbit coupling parameters of the acoustic and optical bands read
\be
\zeta_{\bs k}\pm\eta_{+,\bs k}=\frac{S\mu_0 (g\mu_B)^2}{2} (F_{\bs k}^{d_1 a_1} \mp \frac{F_{\bs k}^{d_1 a_2}+ F_{\bs k}^{d_2 a_1}}{2}),
\ee
which contain nine parameters $\beta^{\alpha\beta}_{ij}$. However, these parameters are actually not independent, because any vector ${\bs R}^{d_2a_1}$ (${\bs R}^{d_1a_2}$) can be obtained by rotating one particular ${\bs R}^{d_1a_1}$ around $z$-axis counterclockwise by $2\pi/3$ ($-2\pi/3$). Specifically, we find
\ber
    {\bs R}_{mn,\|}^{d_1a_1}&=&
a_0\left(\begin{array}{c}
\sqrt{3}m-\frac{\sqrt{3}}{2}n-\frac{2\sqrt{3}}{3}\\
\frac{3}{2}n
\end{array}\right),\\
{\bs R}^{d_2a_1}_{(m-n)(-n),\|}&=&
a_0\left(\begin{array}{c}
-\frac{\sqrt{3}}{2}m-\frac{\sqrt{3}}{2}n+\frac{\sqrt{3}}{3}\\
\frac{3}{2}m-\frac{3}{2}n-1
\end{array}\right)\nonumber\\
&=&{\cal R}(2\pi/3){\bs R}^{d_1 a_1}_{mn,\|},\\
{\bs R}^{d_1a_2}_{(-n)(m-4n-2),\|}&=&
a_0\left(\begin{array}{c}
-\frac{\sqrt{3}}{2}m+\sqrt{3}n+\frac{\sqrt{3}}{3}\\
-\frac{3}{2}m+1
\end{array}\right)\nonumber\\
&=&{\cal R}(-2\pi/3){\bs R}^{d_1 a_1}_{mn,\|},
\eer
where the rotation operator around normal direction is defined as
\be
   {\cal R}(\phi)	=	\left(\begin{array}{cc}
\cos\phi & -\sin\phi\\
\sin\phi & \cos\phi
\end{array}\right).
   \ee

By writing the $\beta^{\alpha\beta}_{ij}$ in the form of matrices, we express them as
\ber
\hat{\beta}^{d_1a_1}&=&\left(\begin{array}{cc}
\beta_{xx} & \beta_{xy}\\
\beta_{xy} & \beta_{yy}
\end{array}\right),\\
\hat{\beta}^{d_2 a_1}&=&{\cal R}(2\pi/3)\hat{\beta}(\zeta){\cal R}(-2\pi/3),\\
\hat{\beta}^{d_1a_2}&=&{\cal R}(-2\pi/3)\hat{\beta}(\zeta){\cal R}(2\pi/3),
\eer
which give
 \ber
 F_{\bs k}^{d_1 a_1}&=&\beta_{yy}-\beta_{xx}+2i\beta_{xy}+2\pi ke^{-2i\theta_{\boldsymbol{k}}},\\
 F_{\bs k}^{d_2 a_1}&=&(\frac{1}{2}-\frac{i\sqrt{3}}{2})(\beta_{xx}-\beta_{yy})\nonumber\\
 &&-2(\frac{\sqrt{3}}{2}+\frac{i}{2})\beta_{xy}+2\pi ke^{-2i\theta_{\boldsymbol{k}}},\\
 F_{\bs k}^{d_1 a_2}&=&(\frac{1}{2}+\frac{i\sqrt{3}}{2})(\beta_{xx}-\beta_{yy})\nonumber\\
 &&-2(-\frac{\sqrt{3}}{2}+\frac{i}{2})\beta_{xy}+2\pi ke^{-2i\theta_{\boldsymbol{k}}}.
\eer
In addition, for any vector ${\bs R}_{mn}^{a_1d_1}=(X_{mn}^{a_1d_1},Y_{mn}^{a_1d_1},Z_{mn}^{a_1d_1})=(\sqrt 3 m -\sqrt 3 n/2-2\sqrt 3/3,3n/2,2\eta_z)a_0$ with nonzero $n$, one can always find another vector ${\bs R}_{(m-n) (-n)}^{a_1d_1}=(\sqrt 3 m-\sqrt 3 n/2-2\sqrt 3/3,-3n/2,2\eta_z)a_0=(X_{mn}^{a_1d_1},-Y_{mn}^{a_1d_1},,Z_{mn}^{a_1d_1})$. Their contributions to $\beta_{xy}$ cancel with each other, meaning $\beta_{xy}\equiv 0$ in the present lattice.

Finally, we obtain
\ber
\zeta_{\bs k}\pm {\eta}_{+,\bs k}&=&(1\pm\frac{1}{2})\mu_{0}\mu_{B}M_{s}^{{\rm 2D}}(\beta_{yy}-\beta_{xx})\nonumber\\
&&+(1\mp 1)\mu_{0}\mu_{B}M_{s}^{{\rm 2D}}2\pi ke^{-2i\theta_{\boldsymbol{k}}}
\eer
with $M_{s}^{{\rm 2D}}=Sg^2\mu_{B}/(2A)$. Thus, the spin-orbit coupling in the acoustic and optical bands become
\be
\zeta_{\bs k}-{\eta}_{+,\bs k}=-f_{0}+v_{0}ke^{-2i\theta_{\boldsymbol{k}}},
\ee
and
\be
\zeta_{\bs k}+{\eta}_{+,\bs k}=-3f_{0},
\ee
respectively. Here, $f_{0}=\frac{1}{2}\mu_{0}\mu_{B}M_{s}^{{\rm 2D}}(\beta_{xx}-\beta_{yy})$ and $v_{0}=\mu_{0}\mu_{B}4\pi M_{s}^{{\rm 2D}}$. The value of $f_0$ can be determined from the numerical evaluation at $k=0$ in a lattice model.

Similarly, for the intralayer parameters, we have
\be
G^{\alpha\beta}_{\bs k}\simeq-\frac{1}{A}\beta_{zz}^{\alpha\beta}-\frac{1}{A}\int_{r>\rho}d\boldsymbol{r}e^{i{\bs k}\cdot\boldsymbol{r}}\partial_{z}^{2}\frac{1}{\sqrt {r^2+h^2}}.
\ee
Using $\sqrt{r^2+h^2}\approx r$ for $r\ge \rho$, the second term gives
\ber
\int_{r>\rho}dxdye^{i\boldsymbol{k}\cdot\boldsymbol{r}}\partial_{z}^{2}\frac{1}{\sqrt{r^{2}+h^{2}}}
&\simeq& k\int_{0}^{2\pi}d\theta\int_{k\rho}^{\infty}d\xi e^{i\xi\cos\theta}\nonumber\\
&&=2\pi k,
\eer
leading to
\ber
\epsilon_{\bs k}'&=&A_{\bs k}^{a_1a_1}=-\frac{S\mu_0 (g\mu_B)^2}{2} G^{a_1a_1}_{\bs k}\nonumber\\
&=&\mu_{0}\mu_{B}M_{s}^{{\rm 2D}}(\beta_{zz}^{a_1a_1}+2\pi k)=f_z+\frac{1}{2}v_0 k,\\
\lambda_{\bs k}'&=&A_{\bs k}^{a_1a_2}=-\frac{S\mu_0 (g\mu_B)^2}{2} G^{a_1a_2}_{\bs k}\nonumber\\
&=&\mu_{0}\mu_{B}M_{s}^{{\rm 2D}}(\beta_{zz}^{a_1a_2}+2\pi k)=f_z'+\frac{1}{2} v_0 k.
\eer

\section{Relation between the spin-orbit coupling parameters at $K$ and $K'$ points} \label{socKKm}
According to the rotation relation addressed above, one express the vectors as
\ber
    {\bs R}_{mn,\|}^{d_1a_1}
&=&\left(\begin{array}{c}
     R_{mn,\|}^{d_1a_1}\cos\theta_{mn}\\
     R_{mn,\|}^{d_1a_1}\sin\theta_{mn}
   \end{array}\right),\\
{\bs R}^{d_2a_1}_{(m-n)(-n),\|}
&=&\left(\begin{array}{c}
     R_{mn,\|}^{d_1a_1}\cos(\theta_{mn}+\frac{2\pi}{3})\\
     R_{mn,\|}^{d_1a_1}\sin(\theta_{mn}+\frac{2\pi}{3})
   \end{array}\right),\\
{\bs R}^{d_1a_2}_{(-n)(m-4n-2),\|}
&=&\left(\begin{array}{c}
     R_{mn,\|}^{d_1a_1}\cos(\theta_{mn}-\frac{2\pi}{3})\\
     R_{mn,\|}^{d_1a_1}\sin(\theta_{mn}-\frac{2\pi}{3})
   \end{array}\right).
\eer
By applying them into the general expression of spin-orbit coupling parameters, we have
\ber
\zeta_{\boldsymbol{K}(\boldsymbol{K}')}&=&-\frac{3S\mu_{0}(g\mu_{B})^{2}}{2}\sum_{mn}\frac{(R_{mn,\|}^{d_1a_1})^2}{(R_{mn}^{d_1 a_1})^{5}}e^{-2i\theta_{mn}}\nonumber\\
&&\hspace{2.5cm}\times e^{i\boldsymbol{K}(\boldsymbol{K}')\cdot\boldsymbol{R}_{mn}^{d_1 a_1}}\\
\eta^\prime_{\boldsymbol{K}(\boldsymbol{K}')}&=&-\frac{3S\mu_{0}(g\mu_{B})^{2}}{2}\sum_{mn}\frac{(R_{mn,\|}^{d_1a_1})^2}{(R_{mn}^{d_1 a_1})^{5}}e^{-2i(\theta_{mn}-\frac{2\pi}{3})}\nonumber\\
&&\hspace{2.5cm}\times e^{i\boldsymbol{K}(\boldsymbol{K}')\cdot\boldsymbol{R}_{(-n)(m-4n-2)}^{d_1a_2}}\\
\eta_{\boldsymbol{K}(\boldsymbol{K}')}&=&-\frac{3S\mu_{0}(g\mu_{B})^{2}}{2}\sum_{mn}\frac{(R_{mn,\|}^{d_1a_1})^2}{(R_{mn}^{d_1 a_1})^{5}}e^{-2i(\theta_{mn}+\frac{2\pi}{3})}\nonumber\\
&&\hspace{2.5cm} \times e^{i\boldsymbol{K}(\boldsymbol{K}')\cdot\boldsymbol{R}_{(m-n)(-n)}^{d_2a_1}}
\eer
at $K$ and $K'$ points. By further substituting
\ber
\boldsymbol{K}(\bs K')&=&(\mp\frac{4\pi}{3\sqrt{3}a_0},0),
\eer
we obtain the relations
\ber
\zeta_{\boldsymbol{K}}&=&\eta^\prime_{\boldsymbol{K}}=\eta_{\boldsymbol{K}}e^{-i4\pi/3},\\
\zeta_{\boldsymbol{K}'}&=&\eta_{\boldsymbol{K}'}=\eta^\prime_{\boldsymbol{K}'}e^{i4\pi/3}.
\eer
Here, the upper and lower signs stand for the $K$ and $K'$, respectively. 

For the parameter $\lambda_{\bf k}'$, the involved vectors satisfy $C_3$ rotation symmetry, i.e.,
\ber
\boldsymbol{R}_{mn,\|}^{a_{1}a_{2}}&=&a_{0}\left(\begin{array}{c}
\sqrt{3}m+\frac{\sqrt{3}}{2}n\\
\frac{3}{2}n+1
\end{array}\right),\label{Ra1a2}\\
\boldsymbol{R}_{(-m-n)(m-1),\|}^{a_{1}a_{2}}&=&
{\cal R}(2\pi/3)\boldsymbol{R}_{mn,\|}^{a_{1}a_{2}},\\
\boldsymbol{R}_{(n+1)(-m-n-1),\|}^{a_{1}a_{2}}&=&{\cal R}(-2\pi/3)\boldsymbol{R}_{m,\|}^{a_{1}a_{2}}.\label{Ra1a2b}
\eer
This allows us to transform the summation around $K$ ($K'$) into
\ber
\lambda_{\boldsymbol{K}(\boldsymbol{K}')+\boldsymbol{q}}^{\prime}&=&-\frac{S\mu_{0}(g\mu_{B})^{2}}{2}\sum_{mn}\frac{(R_{mn}^{a_{1}a_{2}})-3(Z_{mn}^{a_{1}a_{2}})^{2}}{(R_{mn}^{a_1a_2})^{5}}\nonumber\\
&&\hspace{2cm}\times e^{i(\mp\mathbf{K}_{0}+\boldsymbol{q})\cdot\boldsymbol{R}_{mn}^{a_{1}a_{2}}}\\
&=&-\frac{S\mu_{0}(g\mu_{B})^{2}}{2}\sum_{mn}\frac{(R_{mn}^{a_{1}a_{2}})-3(Z_{mn}^{a_{1}a_{2}})^{2}}{(R_{mn}^{a_1a_2})^{5}}\nonumber\\
&&\hspace{-1cm}\times\frac{1}{3}[e^{i(\mp\mathbf{K}_{0}+\boldsymbol{q})\cdot\boldsymbol{R}_{mn}^{a_{1}a_{2}}}+e^{i(\mp\mathbf{K}_{0}+\boldsymbol{q})\cdot\boldsymbol{R}_{(-m-n)(m-1)}^{a_{1}a_{2}}}\nonumber\\
  &&+e^{i(\mp\mathbf{K}_{0}+\boldsymbol{q})\cdot\boldsymbol{R}_{(n+1)(-m-n-1)}^{a_{1}a_{2}}}].
\eer
We then expand it up to the linear order in $q$. Actually the zero-th order vanishes. The linear order leads to
\be
\lambda_{\boldsymbol{K}(\boldsymbol{K}')+\boldsymbol{q}}^{\prime}=\pm e^{\pm i\theta_{\boldsymbol{q}}}qv_{m}\label{A12}
\ee
with
\ber
v_{m}&=&-\frac{S\mu_{0}(g\mu_{B})^{2}}{4}\sum_{mn}\frac{(R_{mn}^{a_{1}a_{2}})-3(Z_{mn}^{a_{1}a_{2}})^{2}}{(R_{mn}^{a_1a_2})^{4}}\nonumber\\
  &&\hspace{ 2cm}\times\sin[\frac{4\pi}{3}(m-n)+\theta_{mn}].
\eer
Eq.~(\ref{A12}) has the same form as the exchange term
\be
\lambda_{\boldsymbol{K}(\boldsymbol{K}')+\boldsymbol{q}}=-\omega_{\rm ex}\gamma_{\boldsymbol{K}(\boldsymbol{K}')+\boldsymbol{q}}=\mp v_{F}qe^{\pm i\theta_{\boldsymbol{q}}},
\ee
where the exchange-induced velocity $v_{F}=a_{0}\omega_{{\rm ex}}/2$. As a result, we obtain
\be
\lambda_{\boldsymbol{K}(\boldsymbol{K}')+\boldsymbol{q}}+\lambda_{\boldsymbol{K}(\boldsymbol{K}')+\boldsymbol{q}}^{\prime}=\mp(v_{F}-v_{m})qe^{\pm i\theta_{\boldsymbol{q}}}.
\ee
Typically, the intralayer exchange interaction is much stronger than DDI, which leads to $v_F\gg v_m$ and 
\ber
\eta_{\pm,\boldsymbol{K}+\boldsymbol{q}}&=&-(\eta_{\boldsymbol{K}}^{\prime}e^{-i\theta_{\boldsymbol{q}}}\pm\eta_{\boldsymbol{K}}e^{i\theta_{\boldsymbol{q}}})/2\nonumber\\
&=&-\zeta_{\boldsymbol{K}}[e^{-i\theta_{\boldsymbol{q}}}\pm e^{i(\theta_{\boldsymbol{q}}-2\pi/3)}]/2,\\
\eta_{\pm,\boldsymbol{K}'+\boldsymbol{q}}&=&(\eta_{\boldsymbol{K}'}^{\prime}e^{i\theta_{\boldsymbol{q}}}\pm\eta_{\boldsymbol{K}'}e^{-i\theta_{\boldsymbol{q}}})/2\nonumber\\
&=&\pm\zeta_{\boldsymbol{K}'}[e^{-i\theta_{\boldsymbol{q}}}\pm e^{i(\theta_{\boldsymbol{q}}+2\pi/3)}]/2.
\eer

By using Eqs.~(\ref{Ra1a2})-(\ref{Ra1a2b}), one can also calculate the particle-hole coupling induced by intralayer DDI 
\ber
B_{\boldsymbol{K}(\boldsymbol{K}')}^{a_{1}a_{2}}&=&B_{\boldsymbol{K}'(\boldsymbol{K})}^{a_{2}a_{1}}\nonumber \\
&=&-\frac{3S\mu_{0}(g\mu_{B})^{2}}{2}\sum_{mn}\frac{e^{-2i\theta_{mn}}}{(R_{mn}^{a_1a_2})^{3}}e^{\mp i\mathbf{K}_{0}\cdot\boldsymbol{R}_{mn}^{a_{1}a_{2}}}\nonumber \\
&=&-\frac{S\mu_{0}(g\mu_{B})^{2}}{2}\sum_{mn}\frac{e^{-2i\theta_{mn}}}{(R_{mn}^{a_{1}a_{2}})^{3}}e^{\mp i\mathbf{K}_{0}\cdot\boldsymbol{R}_{mn}^{a_{1}a_{2}}}\nonumber\\
&&+\frac{e^{-2i\theta_{(-m-n)(m-1)}}}{(R_{(-m-n)(m-1)}^{a_{1}a_{2}})^{3}}e^{\mp i\mathbf{K}_{0}\cdot\boldsymbol{R}_{(-m-n)(m-1)}^{a_{1}a_{2}}}\nonumber\\
&&+\frac{e^{-2i\theta_{(n+1)(-m-n-1)}}}{(R_{(n+1)(-m-n-1)}^{a_{1}a_{2}})^{3}}e^{\mp i\mathbf{K}_{0}\cdot\boldsymbol{R}_{(n+1)(-m-n-1)}^{a_{1}a_{2}}}\nonumber \\
&=&f_K [1+e^{i\frac{2\pi}{3}(1\pm1)}+e^{-i\frac{2\pi}{3}(1\pm1)}]\label{Ba1a2K}.
\eer
with
\be
f_K=-\frac{S\mu_{0}(g\mu_{B})^{2}}{2}\sum_{mn}\frac{e^{-2i\theta_{mn}}}{(R_{mn}^{a_1a_2})^{3}}e^{\mp i\frac{4\pi}{3}(m-n)}.
\ee

\bibliography{Refs}

\begin{thebibliography}{28}
\expandafter\ifx\csname natexlab\endcsname\relax\def\natexlab#1{#1}\fi
\expandafter\ifx\csname bibnamefont\endcsname\relax
  \def\bibnamefont#1{#1}\fi
\expandafter\ifx\csname bibfnamefont\endcsname\relax
  \def\bibfnamefont#1{#1}\fi
\expandafter\ifx\csname citenamefont\endcsname\relax
  \def\citenamefont#1{#1}\fi
\expandafter\ifx\csname url\endcsname\relax
  \def\url#1{\texttt{#1}}\fi
\expandafter\ifx\csname urlprefix\endcsname\relax\def\urlprefix{URL }\fi
\providecommand{\bibinfo}[2]{#2}
\providecommand{\eprint}[2][]{\url{#2}}

\bibitem[{\citenamefont{Gong et~al.}(2017)\citenamefont{Gong, Li, Li, Ji,
  Stern, Xia, Cao, Bao, Wang, Wang et~al.}}]{xZhang17}
\bibinfo{author}{\bibfnamefont{C.}~\bibnamefont{Gong}},
  \bibinfo{author}{\bibfnamefont{L.}~\bibnamefont{Li}},
  \bibinfo{author}{\bibfnamefont{Z.}~\bibnamefont{Li}},
  \bibinfo{author}{\bibfnamefont{H.}~\bibnamefont{Ji}},
  \bibinfo{author}{\bibfnamefont{A.}~\bibnamefont{Stern}},
  \bibinfo{author}{\bibfnamefont{Y.}~\bibnamefont{Xia}},
  \bibinfo{author}{\bibfnamefont{T.}~\bibnamefont{Cao}},
  \bibinfo{author}{\bibfnamefont{W.}~\bibnamefont{Bao}},
  \bibinfo{author}{\bibfnamefont{C.}~\bibnamefont{Wang}},
  \bibinfo{author}{\bibfnamefont{Y.}~\bibnamefont{Wang}}, \bibnamefont{et~al.},
  \bibinfo{journal}{Nature} \textbf{\bibinfo{volume}{546}},
  \bibinfo{pages}{265} (\bibinfo{year}{2017}),
  \urlprefix\url{https://www.nature.com/articles/nature22060}.

\bibitem[{\citenamefont{Huang et~al.}(2017)\citenamefont{Huang, Clark,
  Navarro-Moratalla, Klein, Cheng, Seyler, Zhong, Schmidgall, McGuire, Cobden
  et~al.}}]{xdXu17}
\bibinfo{author}{\bibfnamefont{B.}~\bibnamefont{Huang}},
  \bibinfo{author}{\bibfnamefont{G.}~\bibnamefont{Clark}},
  \bibinfo{author}{\bibfnamefont{E.}~\bibnamefont{Navarro-Moratalla}},
  \bibinfo{author}{\bibfnamefont{D.~R.} \bibnamefont{Klein}},
  \bibinfo{author}{\bibfnamefont{R.}~\bibnamefont{Cheng}},
  \bibinfo{author}{\bibfnamefont{K.~L.} \bibnamefont{Seyler}},
  \bibinfo{author}{\bibfnamefont{D.}~\bibnamefont{Zhong}},
  \bibinfo{author}{\bibfnamefont{E.}~\bibnamefont{Schmidgall}},
  \bibinfo{author}{\bibfnamefont{M.~A.} \bibnamefont{McGuire}},
  \bibinfo{author}{\bibfnamefont{D.~H.} \bibnamefont{Cobden}},
  \bibnamefont{et~al.}, \bibinfo{journal}{Nature}
  \textbf{\bibinfo{volume}{546}}, \bibinfo{pages}{270} (\bibinfo{year}{2017}),
  \urlprefix\url{https://www.nature.com/articles/nature22391}.

\bibitem[{\citenamefont{Mermin and Wagner}(1966)}]{Mermin66}
\bibinfo{author}{\bibfnamefont{N.~D.} \bibnamefont{Mermin}} \bibnamefont{and}
  \bibinfo{author}{\bibfnamefont{H.}~\bibnamefont{Wagner}},
  \bibinfo{journal}{Phys. Rev. Lett.} \textbf{\bibinfo{volume}{17}},
  \bibinfo{pages}{1133} (\bibinfo{year}{1966}),
  \urlprefix\url{https://link.aps.org/doi/10.1103/PhysRevLett.17.1133}.

\bibitem[{\citenamefont{Sivadas et~al.}(2018)\citenamefont{Sivadas, Okamoto,
  Xu, Fennie, and Xiao}}]{Dixiao18}
\bibinfo{author}{\bibfnamefont{N.}~\bibnamefont{Sivadas}},
  \bibinfo{author}{\bibfnamefont{S.}~\bibnamefont{Okamoto}},
  \bibinfo{author}{\bibfnamefont{X.}~\bibnamefont{Xu}},
  \bibinfo{author}{\bibfnamefont{C.~J.} \bibnamefont{Fennie}},
  \bibnamefont{and} \bibinfo{author}{\bibfnamefont{D.}~\bibnamefont{Xiao}},
  \bibinfo{journal}{Nano Letters} \textbf{\bibinfo{volume}{18}},
  \bibinfo{pages}{7658} (\bibinfo{year}{2018}),
  \urlprefix\url{https://doi.org/10.1021/acs.nanolett.8b03321}.

\bibitem[{\citenamefont{Jiang et~al.}(2019)\citenamefont{Jiang, Wang, Chen,
  Zhong, Yuan, Lu, and Ji}}]{Jiwei19}
\bibinfo{author}{\bibfnamefont{P.}~\bibnamefont{Jiang}},
  \bibinfo{author}{\bibfnamefont{C.}~\bibnamefont{Wang}},
  \bibinfo{author}{\bibfnamefont{D.}~\bibnamefont{Chen}},
  \bibinfo{author}{\bibfnamefont{Z.}~\bibnamefont{Zhong}},
  \bibinfo{author}{\bibfnamefont{Z.}~\bibnamefont{Yuan}},
  \bibinfo{author}{\bibfnamefont{Z.-Y.} \bibnamefont{Lu}}, \bibnamefont{and}
  \bibinfo{author}{\bibfnamefont{W.}~\bibnamefont{Ji}}, \bibinfo{journal}{Phys.
  Rev. B} \textbf{\bibinfo{volume}{99}}, \bibinfo{pages}{144401}
  (\bibinfo{year}{2019}),
  \urlprefix\url{https://link.aps.org/doi/10.1103/PhysRevB.99.144401}.

\bibitem[{\citenamefont{Sun et~al.}(2019)\citenamefont{Sun, Yi, Song, Clark,
  Huang, Shan, Wu, Huang, Gao, Chen et~al.}}]{SWWu19}
\bibinfo{author}{\bibfnamefont{Z.}~\bibnamefont{Sun}},
  \bibinfo{author}{\bibfnamefont{Y.}~\bibnamefont{Yi}},
  \bibinfo{author}{\bibfnamefont{T.}~\bibnamefont{Song}},
  \bibinfo{author}{\bibfnamefont{G.}~\bibnamefont{Clark}},
  \bibinfo{author}{\bibfnamefont{B.}~\bibnamefont{Huang}},
  \bibinfo{author}{\bibfnamefont{Y.}~\bibnamefont{Shan}},
  \bibinfo{author}{\bibfnamefont{S.}~\bibnamefont{Wu}},
  \bibinfo{author}{\bibfnamefont{D.}~\bibnamefont{Huang}},
  \bibinfo{author}{\bibfnamefont{C.}~\bibnamefont{Gao}},
  \bibinfo{author}{\bibfnamefont{Z.}~\bibnamefont{Chen}}, \bibnamefont{et~al.},
  \bibinfo{journal}{Nature} \textbf{\bibinfo{volume}{572}},
  \bibinfo{pages}{497} (\bibinfo{year}{2019}),
  \urlprefix\url{https://www.nature.com/articles/s41586-019-1445-3}.

\bibitem[{\citenamefont{Wang et~al.}(2014)\citenamefont{Wang, Du, Hammel, and
  Yang}}]{WangHL2014}
\bibinfo{author}{\bibfnamefont{H.}~\bibnamefont{Wang}},
  \bibinfo{author}{\bibfnamefont{C.}~\bibnamefont{Du}},
  \bibinfo{author}{\bibfnamefont{P.~C.} \bibnamefont{Hammel}},
  \bibnamefont{and} \bibinfo{author}{\bibfnamefont{F.}~\bibnamefont{Yang}},
  \bibinfo{journal}{Phys. Rev. Lett.} \textbf{\bibinfo{volume}{113}},
  \bibinfo{pages}{097202} (\bibinfo{year}{2014}),
  \urlprefix\url{https://link.aps.org/doi/10.1103/PhysRevLett.113.097202}.

\bibitem[{\citenamefont{Qiu et~al.}(2016)\citenamefont{Qiu, Li, Hou, Arenholz,
  N'Diaye, Tan, ichi Uchida, Sato, Okamoto, Tserkovnyak et~al.}}]{Qiu16}
\bibinfo{author}{\bibfnamefont{Z.}~\bibnamefont{Qiu}},
  \bibinfo{author}{\bibfnamefont{J.}~\bibnamefont{Li}},
  \bibinfo{author}{\bibfnamefont{D.}~\bibnamefont{Hou}},
  \bibinfo{author}{\bibfnamefont{E.}~\bibnamefont{Arenholz}},
  \bibinfo{author}{\bibfnamefont{A.~T.} \bibnamefont{N'Diaye}},
  \bibinfo{author}{\bibfnamefont{A.}~\bibnamefont{Tan}},
  \bibinfo{author}{\bibfnamefont{K.}~\bibnamefont{ichi Uchida}},
  \bibinfo{author}{\bibfnamefont{K.}~\bibnamefont{Sato}},
  \bibinfo{author}{\bibfnamefont{S.}~\bibnamefont{Okamoto}},
  \bibinfo{author}{\bibfnamefont{Y.}~\bibnamefont{Tserkovnyak}},
  \bibnamefont{et~al.}, \bibinfo{journal}{Nature Commun.}
  \textbf{\bibinfo{volume}{7}}, \bibinfo{pages}{12670} (\bibinfo{year}{2016}),
  \urlprefix\url{https://www.nature.com/articles/ncomms12670}.

\bibitem[{\citenamefont{Lebrun et~al.}(2018)\citenamefont{Lebrun, Ross, Bender,
  Qaiumzadeh, Baldrati, Cramer, Brataas, Duine, and Klaui}}]{Lebrun18}
\bibinfo{author}{\bibfnamefont{R.}~\bibnamefont{Lebrun}},
  \bibinfo{author}{\bibfnamefont{A.}~\bibnamefont{Ross}},
  \bibinfo{author}{\bibfnamefont{S.~A.} \bibnamefont{Bender}},
  \bibinfo{author}{\bibfnamefont{A.}~\bibnamefont{Qaiumzadeh}},
  \bibinfo{author}{\bibfnamefont{L.}~\bibnamefont{Baldrati}},
  \bibinfo{author}{\bibfnamefont{J.}~\bibnamefont{Cramer}},
  \bibinfo{author}{\bibfnamefont{A.}~\bibnamefont{Brataas}},
  \bibinfo{author}{\bibfnamefont{R.~A.} \bibnamefont{Duine}}, \bibnamefont{and}
  \bibinfo{author}{\bibfnamefont{M.}~\bibnamefont{Klaui}},
  \bibinfo{journal}{Nature} \textbf{\bibinfo{volume}{561}},
  \bibinfo{pages}{222} (\bibinfo{year}{2018}),
  \urlprefix\url{https://www.nature.com/articles/s41586-018-0490-7}.

\bibitem[{\citenamefont{Li et~al.}(2020)\citenamefont{Li, Wilson, Cheng,
  Lohmann, Kavand, Yuan, Aldosary, Agladze, Wei, Sherwin et~al.}}]{JShi20}
\bibinfo{author}{\bibfnamefont{J.}~\bibnamefont{Li}},
  \bibinfo{author}{\bibfnamefont{C.~B.} \bibnamefont{Wilson}},
  \bibinfo{author}{\bibfnamefont{R.}~\bibnamefont{Cheng}},
  \bibinfo{author}{\bibfnamefont{M.}~\bibnamefont{Lohmann}},
  \bibinfo{author}{\bibfnamefont{M.}~\bibnamefont{Kavand}},
  \bibinfo{author}{\bibfnamefont{W.}~\bibnamefont{Yuan}},
  \bibinfo{author}{\bibfnamefont{M.}~\bibnamefont{Aldosary}},
  \bibinfo{author}{\bibfnamefont{N.}~\bibnamefont{Agladze}},
  \bibinfo{author}{\bibfnamefont{P.}~\bibnamefont{Wei}},
  \bibinfo{author}{\bibfnamefont{M.~S.} \bibnamefont{Sherwin}},
  \bibnamefont{et~al.}, \bibinfo{journal}{Nature}
  \textbf{\bibinfo{volume}{578}}, \bibinfo{pages}{70} (\bibinfo{year}{2020}),
  \urlprefix\url{https://doi.org/10.1038/s41586-020-1950-4}.

\bibitem[{\citenamefont{Vaidya et~al.}(2020)\citenamefont{Vaidya, Morley, van
  Tol, Liu, Cheng, Brataas, Lederman, and del Barco}}]{Vaidya20}
\bibinfo{author}{\bibfnamefont{P.}~\bibnamefont{Vaidya}},
  \bibinfo{author}{\bibfnamefont{S.~A.} \bibnamefont{Morley}},
  \bibinfo{author}{\bibfnamefont{J.}~\bibnamefont{van Tol}},
  \bibinfo{author}{\bibfnamefont{Y.}~\bibnamefont{Liu}},
  \bibinfo{author}{\bibfnamefont{R.}~\bibnamefont{Cheng}},
  \bibinfo{author}{\bibfnamefont{A.}~\bibnamefont{Brataas}},
  \bibinfo{author}{\bibfnamefont{D.}~\bibnamefont{Lederman}}, \bibnamefont{and}
  \bibinfo{author}{\bibfnamefont{E.}~\bibnamefont{del Barco}},
  \bibinfo{journal}{Science} \textbf{\bibinfo{volume}{368}},
  \bibinfo{pages}{160} (\bibinfo{year}{2020}),
  \urlprefix\url{https://science.sciencemag.org/content/368/6487/160.full}.

\bibitem[{\citenamefont{Lan et~al.}(2017)\citenamefont{Lan, Yu, and
  Xiao}}]{Xiao17}
\bibinfo{author}{\bibfnamefont{J.}~\bibnamefont{Lan}},
  \bibinfo{author}{\bibfnamefont{W.}~\bibnamefont{Yu}}, \bibnamefont{and}
  \bibinfo{author}{\bibfnamefont{J.}~\bibnamefont{Xiao}},
  \bibinfo{journal}{Nature Commun.} \textbf{\bibinfo{volume}{8}},
  \bibinfo{pages}{178} (\bibinfo{year}{2017}),
  \urlprefix\url{https://www.nature.com/articles/s41467-017-00265-5}.

\bibitem[{\citenamefont{Shen}(2020)}]{Shen20}
\bibinfo{author}{\bibfnamefont{K.}~\bibnamefont{Shen}}, \bibinfo{journal}{Phys.
  Rev. Lett.} \textbf{\bibinfo{volume}{124}}, \bibinfo{pages}{077201}
  (\bibinfo{year}{2020}),
  \urlprefix\url{https://link.aps.org/doi/10.1103/PhysRevLett.124.077201}.

\bibitem[{\citenamefont{Zhang et~al.}(2020)\citenamefont{Zhang, Li, Weber,
  Goldberger, Mak, and Shan}}]{Zhang20}
\bibinfo{author}{\bibfnamefont{X.-X.} \bibnamefont{Zhang}},
  \bibinfo{author}{\bibfnamefont{L.}~\bibnamefont{Li}},
  \bibinfo{author}{\bibfnamefont{D.}~\bibnamefont{Weber}},
  \bibinfo{author}{\bibfnamefont{J.}~\bibnamefont{Goldberger}},
  \bibinfo{author}{\bibfnamefont{K.~F.} \bibnamefont{Mak}}, \bibnamefont{and}
  \bibinfo{author}{\bibfnamefont{J.}~\bibnamefont{Shan}},
  \bibinfo{journal}{Nature Materials} \textbf{\bibinfo{volume}{19}},
  \bibinfo{pages}{838} (\bibinfo{year}{2020}),
  \urlprefix\url{https://doi.org/10.1038/s41563-020-0713-9}.

\bibitem[{\citenamefont{Cenker et~al.}(2020)\citenamefont{Cenker, Huang, Suri,
  Thijssen, Miller, Song, Taniguchi, Watanabe, McGuire, Xiao
  et~al.}}]{Cenker20}
\bibinfo{author}{\bibfnamefont{J.}~\bibnamefont{Cenker}},
  \bibinfo{author}{\bibfnamefont{B.}~\bibnamefont{Huang}},
  \bibinfo{author}{\bibfnamefont{N.}~\bibnamefont{Suri}},
  \bibinfo{author}{\bibfnamefont{P.}~\bibnamefont{Thijssen}},
  \bibinfo{author}{\bibfnamefont{A.}~\bibnamefont{Miller}},
  \bibinfo{author}{\bibfnamefont{T.}~\bibnamefont{Song}},
  \bibinfo{author}{\bibfnamefont{T.}~\bibnamefont{Taniguchi}},
  \bibinfo{author}{\bibfnamefont{K.}~\bibnamefont{Watanabe}},
  \bibinfo{author}{\bibfnamefont{M.~A.} \bibnamefont{McGuire}},
  \bibinfo{author}{\bibfnamefont{D.}~\bibnamefont{Xiao}}, \bibnamefont{et~al.},
  \emph{\bibinfo{title}{Direct observation of 2d magnons in atomically thin
  cri$_3$}} (\bibinfo{year}{2020}), \bibinfo{note}{arXiv:2001.07025}.

\bibitem[{\citenamefont{Holstein and Primakoff}(1940)}]{Holstein40}
\bibinfo{author}{\bibfnamefont{T.}~\bibnamefont{Holstein}} \bibnamefont{and}
  \bibinfo{author}{\bibfnamefont{H.}~\bibnamefont{Primakoff}},
  \bibinfo{journal}{Phys. Rev.} \textbf{\bibinfo{volume}{58}},
  \bibinfo{pages}{1098} (\bibinfo{year}{1940}),
  \urlprefix\url{https://link.aps.org/doi/10.1103/PhysRev.58.1098}.

\bibitem[{\citenamefont{Shen}(2019)}]{Shen2019c}
\bibinfo{author}{\bibfnamefont{K.}~\bibnamefont{Shen}}, \bibinfo{journal}{Phys.
  Rev. B} \textbf{\bibinfo{volume}{100}}, \bibinfo{pages}{094423}
  (\bibinfo{year}{2019}),
  \urlprefix\url{https://link.aps.org/doi/10.1103/PhysRevB.100.094423}.

\bibitem[{\citenamefont{Chen et~al.}(2018)\citenamefont{Chen, Chung, Gao, Chen,
  Stone, Kolesnikov, Huang, and Dai}}]{LChen18}
\bibinfo{author}{\bibfnamefont{L.}~\bibnamefont{Chen}},
  \bibinfo{author}{\bibfnamefont{J.-H.} \bibnamefont{Chung}},
  \bibinfo{author}{\bibfnamefont{B.}~\bibnamefont{Gao}},
  \bibinfo{author}{\bibfnamefont{T.}~\bibnamefont{Chen}},
  \bibinfo{author}{\bibfnamefont{M.~B.} \bibnamefont{Stone}},
  \bibinfo{author}{\bibfnamefont{A.~I.} \bibnamefont{Kolesnikov}},
  \bibinfo{author}{\bibfnamefont{Q.}~\bibnamefont{Huang}}, \bibnamefont{and}
  \bibinfo{author}{\bibfnamefont{P.}~\bibnamefont{Dai}},
  \bibinfo{journal}{Phys. Rev. X} \textbf{\bibinfo{volume}{8}},
  \bibinfo{pages}{041028} (\bibinfo{year}{2018}),
  \urlprefix\url{https://link.aps.org/doi/10.1103/PhysRevX.8.041028}.

\bibitem[{\citenamefont{You et~al.}(2019)\citenamefont{You, Chen, Zhang, Sheng,
  Yang, and Su}}]{GSu19}
\bibinfo{author}{\bibfnamefont{J.-Y.} \bibnamefont{You}},
  \bibinfo{author}{\bibfnamefont{C.}~\bibnamefont{Chen}},
  \bibinfo{author}{\bibfnamefont{Z.}~\bibnamefont{Zhang}},
  \bibinfo{author}{\bibfnamefont{X.-L.} \bibnamefont{Sheng}},
  \bibinfo{author}{\bibfnamefont{S.~A.} \bibnamefont{Yang}}, \bibnamefont{and}
  \bibinfo{author}{\bibfnamefont{G.}~\bibnamefont{Su}}, \bibinfo{journal}{Phys.
  Rev. B} \textbf{\bibinfo{volume}{100}}, \bibinfo{pages}{064408}
  (\bibinfo{year}{2019}),
  \urlprefix\url{https://link.aps.org/doi/10.1103/PhysRevB.100.064408}.

\bibitem[{\citenamefont{Liu et~al.}(2020)\citenamefont{Liu, Wang, and
  Shen}}]{JLiu20}
\bibinfo{author}{\bibfnamefont{J.}~\bibnamefont{Liu}},
  \bibinfo{author}{\bibfnamefont{L.}~\bibnamefont{Wang}}, \bibnamefont{and}
  \bibinfo{author}{\bibfnamefont{K.}~\bibnamefont{Shen}},
  \bibinfo{journal}{Phys. Rev. Research} \textbf{\bibinfo{volume}{2}},
  \bibinfo{pages}{023282} (\bibinfo{year}{2020}),
  \urlprefix\url{https://link.aps.org/doi/10.1103/PhysRevResearch.2.023282}.

\bibitem[{\citenamefont{Wan et~al.}(2011)\citenamefont{Wan, Turner, Vishwanath,
  and Savrasov}}]{Wan11}
\bibinfo{author}{\bibfnamefont{X.}~\bibnamefont{Wan}},
  \bibinfo{author}{\bibfnamefont{A.~M.} \bibnamefont{Turner}},
  \bibinfo{author}{\bibfnamefont{A.}~\bibnamefont{Vishwanath}},
  \bibnamefont{and} \bibinfo{author}{\bibfnamefont{S.~Y.}
  \bibnamefont{Savrasov}}, \bibinfo{journal}{Phys. Rev. B}
  \textbf{\bibinfo{volume}{83}}, \bibinfo{pages}{205101}
  (\bibinfo{year}{2011}),
  \urlprefix\url{https://link.aps.org/doi/10.1103/PhysRevB.83.205101}.

\bibitem[{\citenamefont{Burkov and Balents}(2011)}]{Burkov11}
\bibinfo{author}{\bibfnamefont{A.~A.} \bibnamefont{Burkov}} \bibnamefont{and}
  \bibinfo{author}{\bibfnamefont{L.}~\bibnamefont{Balents}},
  \bibinfo{journal}{Phys. Rev. Lett.} \textbf{\bibinfo{volume}{107}},
  \bibinfo{pages}{127205} (\bibinfo{year}{2011}),
  \urlprefix\url{https://link.aps.org/doi/10.1103/PhysRevLett.107.127205}.

\bibitem[{\citenamefont{Xu et~al.}(2015)\citenamefont{Xu, Belopolski, Alidoust,
  Neupane, Bian, Zhang, Sankar, Chang, Yuan, Lee et~al.}}]{Hasan15}
\bibinfo{author}{\bibfnamefont{S.-Y.} \bibnamefont{Xu}},
  \bibinfo{author}{\bibfnamefont{I.}~\bibnamefont{Belopolski}},
  \bibinfo{author}{\bibfnamefont{N.}~\bibnamefont{Alidoust}},
  \bibinfo{author}{\bibfnamefont{M.}~\bibnamefont{Neupane}},
  \bibinfo{author}{\bibfnamefont{G.}~\bibnamefont{Bian}},
  \bibinfo{author}{\bibfnamefont{C.}~\bibnamefont{Zhang}},
  \bibinfo{author}{\bibfnamefont{R.}~\bibnamefont{Sankar}},
  \bibinfo{author}{\bibfnamefont{G.}~\bibnamefont{Chang}},
  \bibinfo{author}{\bibfnamefont{Z.}~\bibnamefont{Yuan}},
  \bibinfo{author}{\bibfnamefont{C.-C.} \bibnamefont{Lee}},
  \bibnamefont{et~al.}, \bibinfo{journal}{Science}
  \textbf{\bibinfo{volume}{349}}, \bibinfo{pages}{613} (\bibinfo{year}{2015}),
  \urlprefix\url{https://science.sciencemag.org/content/349/6248/613}.

\bibitem[{\citenamefont{Lv et~al.}(2015)\citenamefont{Lv, Weng, Fu, Wang, Miao,
  Ma, Richard, Huang, Zhao, Chen et~al.}}]{HDing15}
\bibinfo{author}{\bibfnamefont{B.~Q.} \bibnamefont{Lv}},
  \bibinfo{author}{\bibfnamefont{H.~M.} \bibnamefont{Weng}},
  \bibinfo{author}{\bibfnamefont{B.~B.} \bibnamefont{Fu}},
  \bibinfo{author}{\bibfnamefont{X.~P.} \bibnamefont{Wang}},
  \bibinfo{author}{\bibfnamefont{H.}~\bibnamefont{Miao}},
  \bibinfo{author}{\bibfnamefont{J.}~\bibnamefont{Ma}},
  \bibinfo{author}{\bibfnamefont{P.}~\bibnamefont{Richard}},
  \bibinfo{author}{\bibfnamefont{X.~C.} \bibnamefont{Huang}},
  \bibinfo{author}{\bibfnamefont{L.~X.} \bibnamefont{Zhao}},
  \bibinfo{author}{\bibfnamefont{G.~F.} \bibnamefont{Chen}},
  \bibnamefont{et~al.}, \bibinfo{journal}{Phys. Rev. X}
  \textbf{\bibinfo{volume}{5}}, \bibinfo{pages}{031013} (\bibinfo{year}{2015}),
  \urlprefix\url{https://link.aps.org/doi/10.1103/PhysRevX.5.031013}.

\bibitem[{\citenamefont{Aguilera et~al.}(2020)\citenamefont{Aguilera,
  Jaeschke-Ubiergo, Vidal-Silva, Torres, and Nunez}}]{Aguilera20}
\bibinfo{author}{\bibfnamefont{E.}~\bibnamefont{Aguilera}},
  \bibinfo{author}{\bibfnamefont{R.}~\bibnamefont{Jaeschke-Ubiergo}},
  \bibinfo{author}{\bibfnamefont{N.}~\bibnamefont{Vidal-Silva}},
  \bibinfo{author}{\bibfnamefont{L.~E. F.~F.} \bibnamefont{Torres}},
  \bibnamefont{and} \bibinfo{author}{\bibfnamefont{A.~S.} \bibnamefont{Nunez}},
  \bibinfo{journal}{Phys. Rev. B} \textbf{\bibinfo{volume}{102}},
  \bibinfo{pages}{024409} (\bibinfo{year}{2020}),
  \urlprefix\url{https://link.aps.org/doi/10.1103/PhysRevB.102.024409}.

\bibitem[{\citenamefont{Kitaev}(2006)}]{Kitaev06}
\bibinfo{author}{\bibfnamefont{A.}~\bibnamefont{Kitaev}},
  \bibinfo{journal}{Ann. Phys.} \textbf{\bibinfo{volume}{321}},
  \bibinfo{pages}{2} (\bibinfo{year}{2006}),
  \urlprefix\url{https://www.sciencedirect.com/science/article/pii/S0003491605002381?via%3Dihub}.

\bibitem[{\citenamefont{Xu et~al.}(2018)\citenamefont{Xu, Feng, Xiang, and
  Bellaiche}}]{HXiang18}
\bibinfo{author}{\bibfnamefont{C.}~\bibnamefont{Xu}},
  \bibinfo{author}{\bibfnamefont{J.}~\bibnamefont{Feng}},
  \bibinfo{author}{\bibfnamefont{H.}~\bibnamefont{Xiang}}, \bibnamefont{and}
  \bibinfo{author}{\bibfnamefont{L.}~\bibnamefont{Bellaiche}},
  \bibinfo{journal}{npj Computational Materials} \textbf{\bibinfo{volume}{4}},
  \bibinfo{pages}{57} (\bibinfo{year}{2018}),
  \urlprefix\url{https://www.nature.com/articles/s41524-018-0115-6}.

\bibitem[{\citenamefont{Tang et~al.}(2016)\citenamefont{Tang, Zhou, Xu, and
  Zhang}}]{PXTang16}
\bibinfo{author}{\bibfnamefont{P.}~\bibnamefont{Tang}},
  \bibinfo{author}{\bibfnamefont{Q.}~\bibnamefont{Zhou}},
  \bibinfo{author}{\bibfnamefont{G.}~\bibnamefont{Xu}}, \bibnamefont{and}
  \bibinfo{author}{\bibfnamefont{S.-C.} \bibnamefont{Zhang}},
  \bibinfo{journal}{Nature Phys.} \textbf{\bibinfo{volume}{12}},
  \bibinfo{pages}{1100} (\bibinfo{year}{2016}),
  \urlprefix\url{https://www.nature.com/articles/nphys3839}.

\end{thebibliography}

\end{document}